\documentclass[twocolumn]{aastex7}

\newcommand{\citeg}[1]{\citep[e.g.,][]{#1}}
\newcommand{\citsee}[1]{\citep[see,][]{#1}}
\newcommand{\kms}[0]{{\,km\,s$^{-1}$}}

\newcommand{\degree}[0]{{°}}
\renewcommand{\ion}[2]{\textsc{#1}\,\textsc{#2}}

\graphicspath{{figures/}}
\begin{document}

\title{PHR~J1724-3859: A Bipolar Planetary Nebula in Open Cluster Trumpler~25\footnote{based on observations made with the Southern African Large Telescope (SALT), the ESO Vary Large Telescope (VLT) and the Gemini telescope.}}

\author[0000-0002-8634-4204]{Vasiliki Fragkou}
\affiliation{Universidade Federal do Rio de Janeiro, Observat\'orio do Valongo, Ladeira do Pedro Ant\^onio, 43, Sa\'ude CEP 20080-090 Rio de Janeiro, RJ, Brazil}
\email[]{vfragkou@ov.ufrj.br}  

\author[0000-0002-2062-0173]{Quentin A. Parker}
\affiliation{The Laboratory for Space Research, The University of Hong Kong, \\
Cyberport 4, Pokfulam, Hong Kong}
\email[]{quentinp@hku.hk}

\author[0000-0001-9388-7146]{Denise R. Gon\c{c}alves}
\affiliation{Universidade Federal do Rio de Janeiro, Observat\'orio do Valongo, Ladeira do Pedro Ant\^onio, 43, Sa\'ude CEP 20080-090 Rio de Janeiro, RJ, Brazil}
\email[]{}  

\correspondingauthor{Quentin A. Parker, Vasiliki Fragkou }
\email{quentinp@hku.hk, vfragkou@ov.ufrj.br }

%\author[0000-0001-9388-7146]{Denise R. Gonalves}
%\affiliation{Universidade Federal do Rio de Janeiro, Observat\'orio do Valongo, Ladeira do Pedro Ant\^onio, 43, Sa\'ude CEP 20080-090 Rio de Janeiro, RJ, Brazil}

%\author[ 	
%0000-0002-0620-136X]{A. Cortesi}
%\affiliation{Universidade Federal do Rio de Janeiro, Observat\'orio do Valongo, Ladeira do Pedro Ant\^onio, 43, Sa\'ude CEP 20080-090 Rio de Janeiro, RJ, Brazil}
%\affiliation{Instituto de Física, Universidade Federal do Rio de Janeiro, 21941-972, Rio de Janeiro, RJ, Brazil}

%\author[orcid=0000-0000-0000-0002,gname=Bosque, sname='Sur America']{Forrest Sur Am\'{e}rica} 
%\altaffiliation{Las Campanas Observatory}
%\affiliation{Universidad de Chile, Department of Astronomy}
%\email{fakeemail2@google.com}

%\author[gname=Savannah,sname=Africa]{S. Africa}
%\affiliation{South African Astronomical Observatory}
%\affiliation{University of Cape Town, Department of Astronomy}
%\email{fakeemail3@google.com}

%\collaboration{all}{The Terra Mater collaboration}

\begin{abstract}

Planetary nebulae (PNe) studies are essential for understanding late stellar evolution of low-to-intermediate mass stars. 
PNe in open clusters (OC) are rare but valuable since their study directly links their 
properties to those of their progenitors, something that cannot be achieved for field PNe. Here, we report the 
identification of one more OC-PN association to add to the small sample of, now, five pairs. The physical properties of the host cluster, PN and its central star (CSPN) have been explored using high and 
intermediate-resolution spectral and deep photometric data. The close agreement of the radial 
velocities of the PN and host cluster, together with concordance of reddening and distance, show that the 
PN PHR~J1724-3859 is highly likely to be physically associated with the OC Trumpler~25. Deep photometric data allowed clear identification of the CSPN. We find a CSPN effective temperature of around 250 kK and a nebular kinematic age of 23 kyrs, 
both at the extreme end, like the other members of this small class. The progenitor and final CSPN masses have been estimated to be 5.12$_{-0.15}^{+0.16}$ $M_\sun$ and 0.95$\pm$0.12 $M_\sun$ respectively. These latest results agree with the emerging trend for our other OC-PNe, falling below, but approximately parallel to, the latest initial-to-final-mass relation estimates derived from cluster white dwarfs and has important implications for stellar evolution models. All OC-PNe also possess some other common properties (e.g. all are Type-I PNe and bipolars) to be explored in future studies. 

\end{abstract}

%--- \uat{High Energy astrophysics}{739} --- \uat{Interstellar medium}{847} --- \uat{Stellar astronomy}{1583} --- \uat{Solar physics}{1476}}


\section{Introduction} \label{sec:intro}

Planetary nebulae (PNe) are the final stages of evolution of low-to-intermediate mass stars (1-8~M$_\sun$). 
They consist of the remnant core - on the way to becoming or already a white dwarf (WD) and the previously 
ejected material comprising gas and dust. As such, study of the Galactic PN population provides valuable 
information on the chemical enrichment process in our Galaxy, late stage stellar evolution processes, 
Galactic kinematics (using PNe as point probes) and the role of binary stars in ejecta shaping (PN morphology). 

A notable problem in PNe studies is linking PN properties to the properties of their main sequence (MS) 
progenitor stars. This issue can be more easily evaluated for PNe that are physical members of star clusters. 
PN properties can be estimated from spectrophotometric observations, while the progenitor properties 
(since cluster stars share some common physical characteristics) can be determined from cluster colour-magnitude 
diagrams (CMD) and theoretical isochrones \citep{2012MNRAS.427..127B}. Unfortunately, the sample of identified PNe 
in Galactic star clusters is very small, with only four confirmed in globular clusters 
\citep{1928PASP...40..342P,1989ApJ...338..862G,1997AJ....114.2611J} and four in open clusters 
\citep{2011MNRAS.413.1835P,2019MNRAS.484.3078F,2019NatAs...3..851F,2022ApJ...935L..35F,Fragkou2025}. 
Any new cluster-PN identification and confirmation is a valuable datum to add to this rare sample. 
%Since cluster velocity dispersions are typically very low \citep[$\sim1-3$~{\kms},][] {2000ASPC..198..517M} 
such confirmations require consistent and precise cluster and PN radial velocities to within tight errors/dispersions. 
Accurate cluster radial velocities can often been determined from Gaia data of cluster stars \citep{2022A&A...667A.148G}, 
while for the PN, high resolution spectral data are required, unless their faint, low luminosity central 
stars are identified and fall within Gaia's reach.

In Section~{\ref{sec:OC}} and Section~{\ref{sec:PN}} we briefly describe the open cluster and PN. 
In Section~{\ref{sec:obs}} we describe our observations, while in Section~{\ref{sec:res}} we introduce our 
results. Finally, in Section~{\ref{sec:analysis}} we discuss our findings and in Section~{\ref{sec:conc}} 
we state our conclusions. 

\section{The host cluster Trumpler~25} \label{sec:OC}
The open cluster (OC) Trumpler~25 is a relatively well studied OC of medium richness \citep{1975AJ.....80...11V} 
first detected by \citet{1930LicOB..14..154T} and is approximately 9.5~arcmin in diameter. 
The cluster physical parameters have been explored by multiple authors in the literature: 
\citep{2010NewA...15...61S,2013A&A...558A..53K,2014A&A...564A..79D,2017NewA...51...15M,2017MNRAS.470.3937S,
2018A&A...619A.155S,2018A&A...618A..93C,2019ApJS..245...32L,2020A&A...640A...1C,2020MNRAS.495.2882S,2021A&A...647A..19T,
2021A&A...652A.102H,2021MNRAS.504..356D,2022A&A...660A...4H,2022ApJS..259...19L,2023ApJS..268...46L,2024A&A...686A..42H,
2024AJ....167...12C,2025MNRAS.539.2513A}.  The weighted average of the reported literature cluster parameters 
gives a cluster systemic radial velocity of -26.2 $\pm$2.3\kms if we exclude 
the radial velocities estimated by \citet{2023ApJS..268...46L} and \citet{2021A&A...652A.102H} who use older 
catalogues for cluster members and $\leq$ 5 stars. Using the values found by the same authors, we also find 
proper motions of ${\rm pmRA}=0.3\pm0.4$\,mas\,yr$^{-1}$ and ${\rm pmDEC}=-2.1\pm0.3$\,mas\,yr$^{-1}$, 
a cluster mean parallax of $0.46\pm0.05$\,mas, a distance of $1.8\pm0.1$\,kpc, a reddening $E(B-V)=0.89\pm0.04$ 
(implying an extinction of $A_v=2.75\pm0.13$ using the \citet{1989ApJ...345..245C} extinction law with Rv=3.1), 
an age of $118\pm21$\,Myr and an iron abundance of ${\rm [Fe/H]}=0.28\pm0.04$, which implies a metalicity of 
$Z=0.029\pm0.003$ for Z$_\sun=0.0152$ \citep{2012MNRAS.427..127B}. For literature parameter values with 
no assigned errors, we assume an error of 20~\%. 

\section{The OC PN candidate: faint bipolar PN PHR~J1724-3859} \label{sec:PN}
\begin{figure*} %[height=10cm,width=18cm] %GC3_fornax
    \centering
    \includegraphics[width=1.5\columnwidth]{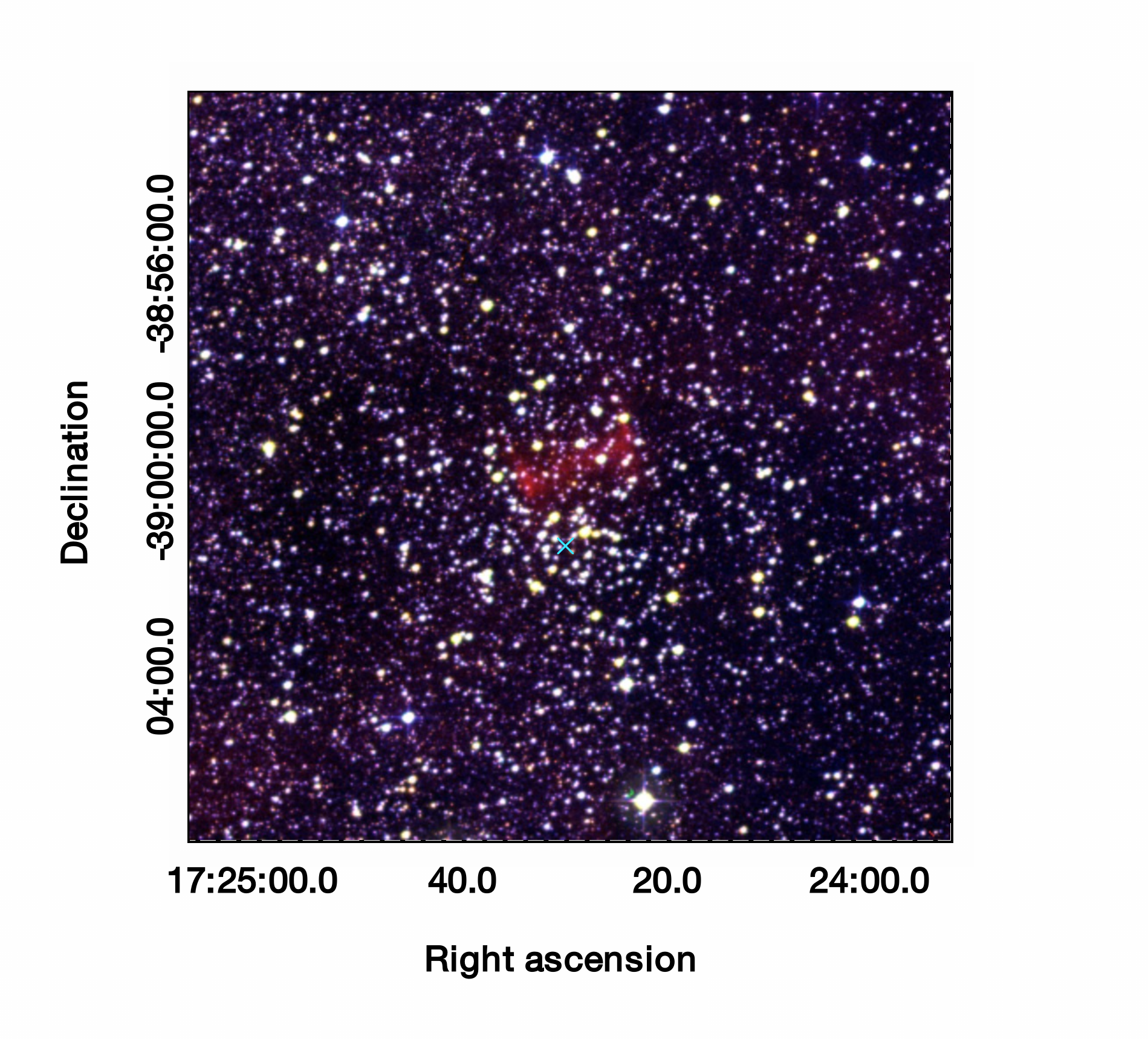}      
    \caption{A 3-colour (R is $H\alpha$, G is short-red and B is blue) SHS image, $15\times15$~arcminutes in size, of the open cluster Trumpler~25 and PN PHR~J1724-3859. North East is to top-left. The image is centred on the faint bipolar PN, clearly identified just 90~arcseconds North of the cluster's reported equatorial coordinate centre of RA:17h24m30.70s and %DEC:-39\degree 01'25" (J2000) that is indicated with the light blue $\times$ marker.}    
} 
    \label{Fig1} 
\end{figure*}

The low surface brightness, extended, bipolar, high excitation PN PHR~J1724-3859 was first 
discovered by visual inspection of the photographic 
SuperCOSMOS H$\alpha$ Survey (SHS) data \citep{2005MNRAS.362..689P} by Parker (private communication) and reported in the 
Macquarie/AAO/Strasbourg H$\alpha$ Planetary Nebula (MASH) catalogue \citep{2006MNRAS.373...79P} and 
subsequently in the Hong Kong/AAO/Strasbourg H$\alpha$ planetary nebula (HASH) database \citep{2016JPhCS.728c2008P} 
where it has HASH ID\#3064. It was considered by us early as an excellent candidate member of OC Trumpler~25 due to its close 
projection on the main cluster area, (see Figure~\ref{Fig1}) and also briefly explored by \citet{2019PhDT........68F}. 
It has an apparent size of $\rm 152\times96~arcsec^2$ 
\citep{2006MNRAS.373...79P}. Preliminary nebular spectra taken in May, 
2000 on the SAAO 1.9~m telescope (PI: Parker) 
show nitrogen enrichments with the observed H$\alpha$/[NII] line ratio being $\sim12$, indicating Type~I 
chemistry \citep{1983IAUS..103..233P}. The [OIII] emission was very weak, likely due to extinction 
\citep{2006MNRAS.373...79P,2019PhDT........68F}. Its integrated H$\alpha$ flux has been estimated to be 
$logF(H\alpha)=-12.17~mW/m^2$ \citep{2013MNRAS.431....2F}. Recently, \citet{2021A&A...656A..51G} reported the 
detection of the central star (CSPN) of PHR~J1724-3859 using Gaia EDR3 data \citep{2022A&A...667A.148G}. 
However, as we show below, this is almost certainly a false identification. Many true CSPN are simply 
too faint to be detected with Gaia where the level of false identification of 
Gaia CSPN reported in \citet{2021A&A...656A..51G} is estimated to be as high as $\sim20\%$ \citep{2022Galax..10...32P}. 
The PN's estimated geometric centre position lies only $\sim90$~arceconds North from Trumpler~25's reported centre 
(see Figure~\ref{Fig1}). This is well within the cluster's tidal radius of $\sim16.6$~arcminutes from the 
physical estimate of 11.6~pc \citet{2024A&A...686A..42H}. 

%\begin{figure}
%\begin{center}
%\includegraphics[width=15cm]{PHR1724-3859e.pdf}   
%\end{center}
%\caption{A 3-colour (R is $H\alpha$, G is short-red and B is blue) SHS image, $15\times15$~arcminutes in size, of the open cluster Trumpler~25 and PN PHR~J1724-3859. North East is to top-left. The image is centred on the faint bipolar PN, clearly identified just 90~arcseconds North of the cluster's reported equatorial coordinate centre of RA:17h24m30.70s and} 

%\label{Fig1}
%\end{figure}

\section{Observations} \label{sec:obs}

Archival narrow-band imaging data (SHS), together with the latest Gaia DR3 \citep{2022A&A...667A.148G} cluster data were used for imaging and photometry. This was supplemented  with our new, high-resolution echelle High Resolution Spectrograph (HRS) data from the 9.2~m South African Large Telescope (SALT) and also with IFU 8.2~m Gemini High-resolution Optical Spectrograph (GHOST) spectral data. We then obtained new, deep cluster photometry with the Focal Reducer and low dispersion Spectrograph (FORS2) mounted on the 8.2~m ESO Very Large Telescope (VLT). We further obtained FORS2 intermediate-resolution long-slit spectra. Combining all these data allowed us to derive new, key PN and cluster properties. The archival SHS data provided initial evidence indicating a likely association of the PN and cluster, while the Gaia DR3 data were used to derive the mean kinematic properties of the cluster (see Subsection~\ref{subsec:T25}) and establish high cluster membership probability supported by our new high-resolution data (see Subsection~\ref{subsec:PHR}). The FORS2 imaging data have been used for the estimation of the cluster mean parameters and the identification of the CSPN (see Subsections~\ref{subsec:T25},~\ref{subsec:CSPN}) and the FORS2 long-slit spectral data allowed a further examination of the nebular spectral properties (see Subsection~\ref{subsec:PHR}).

The aforementioned data show that PN PHR~J1724-3859 is highly likely a physical member of open cluster Trumpler~25. This evaluation is based on the PN and OC radial velocities being closely matched to $\leq3$~{\kms}, within the tight errors (a fundamental requirement), the distances of PN and cluster in excellent concordance, also to within the reasonable errors (another key condition), the reddenings being in close agreement, the PN physical size being plausible as a cluster member and the PN being situated well within the projected cluster tidal radius.

While each compatibility on its own is equivocal, taken together they present a compelling argument for true 
OC-PN association. Indeed, on comparing this example with the four other OC-PN pairs currently known, 
several shared properties, unusual for most PNe, are also found. This further strengthens the arguments (see later). 

\subsection{Intermediate-resolution ESO~VLT FORS2 imaging and spectroscopy}
As a valuable supplement to our large-telescope high resolution spectroscopy from SALT and Gemini and to 
facilitate a more in-depth cluster and PN study, we had earlier collected deep photometric imaging and lower 
resolution spectroscop~ic data for the PN and cluster using the FORS2 imager and spectrograph on the Cassegrain 
focus of the ESO~8.2~m~VLT UT1 telescope. Data were obtained under Program IDs 0103.D-0093(A) and 0103.D0093(B) 
for observations with the E2V (acquisition date: 31 May, 2019) and MIT (acquisition date: 02 June, 2019) 
detectors respectively. 

\subsubsection{FORS2 photometric data}
For an independent study of the cluster physical parameters, and to robustly identify and then 
characterize the CSPN, we collected imaging data with the u-high, b-high, v-high, Bessel R 'special' 
and Bessel I filters with the FORS2 imager centred on the PN. These filter pass-bands are effectively 
equivalent to the UBVRI Johnson-Cousins photometric system \citep{1990PASP..102.1181B,2005ARA&A..43..293B}. 
The FORS2 imager on the VLT has two field of view options, selected via different collimators. In the case of our exposures with the u-high, b-high, v-high and Bessel R special filters, we used the 
standard resolution (SR) collimator with a field of view of $7.1\times7.1$~arcmin to sample as much of the cluster 
as possible in a single pointing, corresponding to pixel scales of 0.25''/pixel. In the case of our exposures using the Bessel I filter, we used the high resolution collimator.
The u-high, b-high and v-high imaging filters were used to construct the cluster CMD and for CSPN identification. 
The Bessel R special and Bessel I filters were used for the further characterization of the CSPN. Both short 
and long exposures were acquired with each filter for the purpose of collecting reliable photometry of both 
bright and faint stellar sources in our field.

Observations with the u-high, b-high and v-high filters (centred on J2000 RA: 17h24m28.78s, DEC:-39\degree00'07.10") 
employed the E2V detector and standard resolution (SR) collimator and a low readout and a binning 
of $\rm 2\times2$. Short exposures of 3~seconds were obtained with each of these three filters, while 
long exposures of 40, 15 and 40 seconds were acquired for the u-high, b-high and v-high filters respectively. 
In all cases the seeing was equal or better than 0.7 arcsec.

The MIT detector was used for observations with the Bessel R 'special' and Bessel I filters (centred on J2000
RA: 17h24m29.23s, DEC:-39\degree00'10.20"). We selected a low readout, and the standard resolution (SR) collimator for the Bessel R special observations and the high resolution (HR) collimator for the Bessel I observations,
plus a binning of $\rm 2\times2$ and $\rm 1\times1$ for acquisitions with the Bessel R special and 
Bessel I filters respectively. A 5 and $\rm 3\times120$ sec short and long exposures were acquired with the 
Bessel R special filter, while a 10 and $\rm 5\times120$ sec exposures were acquired with the Bessel I filter. 
The seeing during these observations was around 1 arcsec.

All imaging data were reduced and processed with EsoReflex \citep{2013A&A...559A..96F}, while the source extraction 
and photometry across Trumpler~25 was done with SExtractor within EsoReflex using mag\_auto and the 
corresponding parameter files for FORS2 (fors2\_stetson.sex, fors.param, fors.conv and fors.nnw). Consequently, 
the data were corrected for atmospheric extinction and colour term correction was applied. Zeropoints, 
extinction coefficients and colour terms for each of 
our filters and CCD chips (FORS2 consists of two separate CCD chips that observe simultaneously) were obtained from 
the FORS2 database\footnote{\url{http://archive.eso.org/bin/qc1_cgi?action=qc1_browse_table&table=fors2_photometry}
} for 
the time period of our observations. Since this information is not available for the u-high filter, u-high data were 
calibrated by cross-correlation with the \citet{2010PASP..122.1437P} U magnitudes. The data were processed and 
calibrated separately for each FORS2 CCD chip, long (Bessel R special and Bessel I long exposures were combined) and 
short exposures. Calibrated data from long and short exposures in each filter were selected accordingly, in order to 
exclude saturated stars, and then combined. This resulted in a source catalogue that contains the maximum possible 
stellar magnitude range. Finally, the data from both CCD chips for each filter were combined.  

\subsubsection{FORS2 spectroscopic data}

PN long-slit medium to low spectral data were acquired with the 1200B+97 (E2V detector, R $\approx$ 1420) and 600RI+19 (MIT detector, R $\approx$ 1000) 
grisms of FORS2. These covered the blue (3660-5110 \AA~ with central $\rm \lambda$= 4350 \AA) and red 
(5120-8450 \AA~ with central $\rm \lambda$= 6780 \AA) parts of the optical spectrum respectively. 
In both cases we used the SR collimator, a low readout and a binning of $\rm 2\times2$. The slit length is 
523~arcsec while a 1~arcsec slit width was selected. Acquisitions with the 1200B+97 (J2000 RA:17h24m28.78d, 
DEC:-39\degree00'07.10") and 600RI+19 (J2000 RA:17h24m29.2s, DEC:-39\degree00'10.20") grisms have a slit position 
angle of 65\degree and 68\degree (N=0\degree, E=90\degree) respectively, moderately inclined relative to the nebular bipolar axis, with an angle of about 65 \degree. The exposure times were 
$3\times900$~seconds with the 1200B+97 grism and $3\times600$~seconds with the 600RI+19 grism. For the flux 
calibration of our data, we additionally observed the spectrophotometric standard star LTT~3864 with both grisms.  

The spectral data were reduced and processed with EsoReflex  \citep{2013A&A...559A..96F} and standard IRAF techniques. 
The data from each CCD chip were processed and calibrated separately before being combined for further analysis. 
The nebula is very extended, covering almost the entire FORS2 field. Consequently, and combined with the complications added by the separate FORS2 CCD chips and the crowded field, there is no adequate 
background sky region available to subtract from our science spectra so the sky subtraction is problematic. This limitation particularly affects the Balmer lines, since these are very bright in the residual sky and cannot be sufficiently subtracted from our science spectra.  
At the lower spectral resolutions here all identified Balmer lines suffer from over-subtraction. These data were taken to
cover a wide optical wavelength range to enable detection of as many PN emission lines as possible. 
The blue VLT FORS2 spectra (see Figure~\ref{Fig5}) clearly shows the [OIII], H$\beta$ and, for the first time, 
the high excitation He~II line, enabling its identification as a high excitation PN. 
Any precise Balmer line ratios derived from analysis of these data 
should be taken with caution and as upper limits. The full log of our observations is presented in Table~\ref{Tab1}).

\subsection{High-resolution spectra from SALT and Gemini telescopes}
For determining the precise systemic radial velocity for PN PHR~J1724-3859 we obtained high-resolution 
spectra across selected, representative parts of the PN using the HRS spectrograph on SALT, a 9.2~m telescope in Sutherland, South Africa and the GHOST spectrograph on the 8.2~m Gemini telescope in Chile. To estimate a precise PN radial velocity 
it is important to calculate its value across multiple symmetrically selected PN pointings to 
account for the internal gas motions (typical PNe expansion velocities are $\sim25-30$~{\kms}). 

\begin{figure*}
\plotone{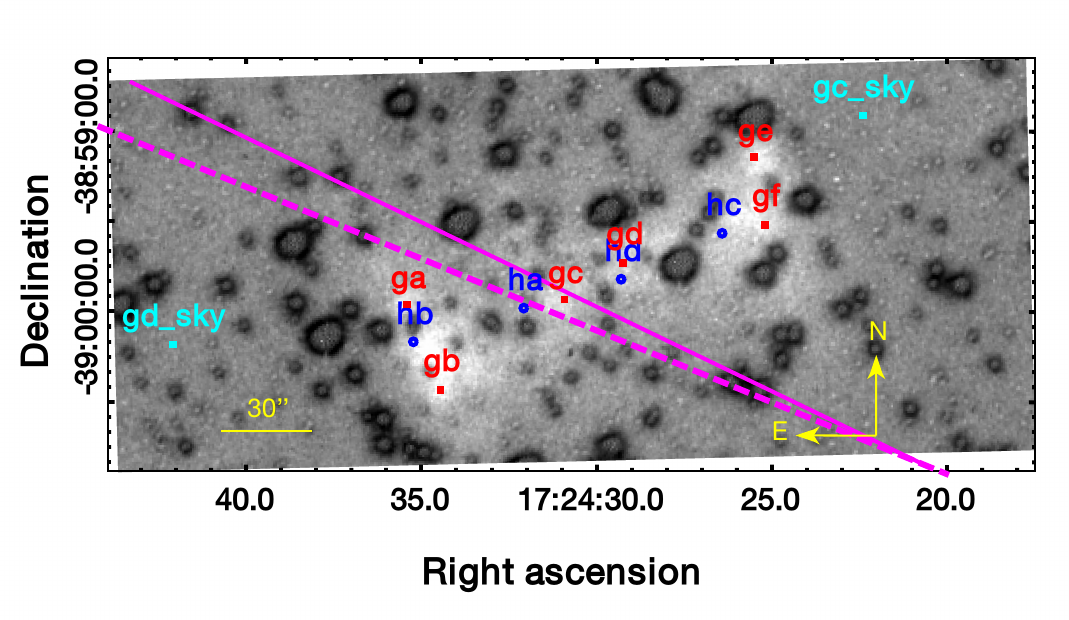}
\caption{A quotient image from the SHS $\rm H\alpha$ divided by the accompanying broad-band `short-red' image of the PN PHR~J1724-3859. The selected PN spectroscopic sampling points for HRS (in blue), GHOST (in red) and sky GHOST (in light blue), together with the FORS2 slits (continuous and dashed magenta lines for the blue and red grisms, respectively) are indicated. The size and shape of the point markings represent the respective HRS and GHOST fields of view which are very similar.}
\label{Fig2}
\end{figure*}

The SALT HRS is a high-resolution dual-beam echelle spectrograph capable of simultaneously acquiring a science 
and a sky frame in the same exposure. The dual beam covers wavelengths 3700-5500\AA~ in the blue and 5500-8900\AA~ 
in the red with a field of view of 2.23~arcsec diameter. Our data were obtained with director's discretionary 
time under program: 2019-1-DDT-002 on the 3rd of June, 2019 in the echelle low-resolution mode with a 
resolution R$\approx$14000. The binning was set to $1\times1$, the 
gain to low while the pixel scale is 0.042 $\AA$/pixel. The seeing during our observations was 1.9~arcsec. We acquired two 900~sec exposures, one for each 
PN bipolar lobe. Due to the nebula extent, the corresponding sky frames are actually additional PN exposures, 
thus giving four PN pointings labelled ha, hb, hc and hd (see Figure~\ref{Fig2}). 
The data have been processed with the standard PySALT/PyHRS\footnote{http://pysalt.salt.ac.za/} 
\citep{2010SPIE.7737E..25C,2016SPIE.9908E..2LC} pipeline.

We also obtained high-resolution spectral data for six additional PN pointings using the GHOST spectrograph 
on the 8.2~m Gemini telescope under program: GS-2024A-Q-232 in June 2024 to further refine the velocities. 
We opted for the dual-IFU (integral field unit) standard-resolution configuration, allowing the 
observation of two separate PN pointings simultaneously. GHOST IFU's have a field of view of 
$1.2\times1.2$\,arcsec$^2$, covering a wavelength range from 3830 to 10000\AA, 
using separate detectors for the blue and the red arm. GHOST's resolution in the standard-resolution mode 
is $R=56000$, allowing a radial velocity precision down to 1~{\kms}. We selected a CCD binning of $1\times2$ 
and a slow readout, while the pixel size in the spatial direction is 0.4~arcsec.

Four $3\times508$~sec exposures allowed the acquisition of spectra in six separate PN pointings (ga-gf), 
plus two sky pointings (gc\_sky, gd\_sky) for the background subtraction (see Figure~\ref{Fig2}). 
All our pointings were carefully selected in areas free of stars to avoid any possible stellar contamination 
in our PN spectra. Pointings ga-gf (observed on the 3rd of July, 2024), gb-ge (observed on the 1st of June, 2024), 
gd-gd\_sky (observed on the 2nd of July, 2024) and gc-gc\_sky (observed on the 3rd of July, 2024) were 
observed simultaneously. The sky pointing gd\_sky was used for the sky subtraction of PN pointings 
ga, gb and gd, while the sky pointing gc\_sky was used for the sky subtraction of PN pointings 
gc, ge and gf. All sky GHOST pointings were observed more than a month after the acquisition of the data for 
PN pointings gb and ge due to scheduling issues. The sky subtraction for these pointings is therefore not optimal, but still sufficient 
for our purposes. The seeing during our June and July observations was $\approx$~0.8 and 0.7 arcsec respectively. 
The airmass during all our observations was around 1.

%\begin{figure*} %[height=10cm,width=18cm] %GC3_fornax
%    \centering
%    \includegraphics[width=1.5\columnwidth]{PHR1724-3859e.pdf}      
%    \caption{A 3-colour (R is $H\alpha$, G is short-red and B is blue) SHS image, $15\times15$~arcminutes in size, of the open cluster Trumpler~25 and PN PHR~J1724-3859. North East is to top-left. The image is centred on the faint bipolar PN, clearly identified just 90~arcseconds North of the cluster's reported equatorial coordinate centre of RA:17h24m30.70s and
%%DEC:-39\degree 01'25" (J2000) that is indicated with the light blue $\times$ marker.}    
%} 
%    \label{Fig1} 
%\end{figure*}

For reducing our HRS and GHOST data we employed the PySALT/PyHRS and DRAGONS 3.20 pipelines respectively, 
and standard IRAF techniques. Standard thorium-argon arc lamps, acquired before and after the science 
exposures, were used for the wavelength 
calibration in each case. GHOST is primarily intended to observe point sources and our data 
have no continuum, thus DRAGONS cannot handle the 1-D spectra extraction automatically. 
For extracting the 1-D spectra we had to make some additional adjustments following \citet{Fragkou2025}. 
Flux calibration was not applied since it is not required for our radial velocity purposes, while barycentric 
correction was turned on. The sky frames were subtracted from the science ones with the IRAF task {\sc images/imarith}.

\begin{table*}[]
\caption{The complete log of our observations.}
\label{Tab1}
\begin{tabular}{lcclcclcc}
\hline \hline
ID        & RA      & DEC     & Instrument  & Mode  & R\footnote{Resolving power.}    & Date & Exp. time  & Seeing    \\
       &           &           &                                     &      & &  & (sec) & ('')    \\
\hline
ha        & 17:24:32 & -38:59:59 & HRS/SALT     & low resolution\footnote{Refers to the instrument mode used, not to be confused with the actual resolution of the spectrograph.}                        &   14000 &                03/06/19       & 900                 &  $\sim$1.9 \\
hb        & 17:24:35 & -39:00:10 & HRS/SALT     & low resolution                        &     14000 &                 03/06/19      & 900                 & $\sim$1.9 \\
hc        & 17:24:26 & -38:59:34 & HRS/SALT     & low resolution                        &    14000 &                 03/06/19      & 900                 &  $\sim$1.9 \\
hd        & 17:24:29 & -38:59:49 & HRS/SALT     & low resolution  &       14000 &                                    03/06/19       & 900                 & $\sim$1.9 \\
ga        & 17:24:35 & -38:59:58 & GHOST/Gemini & double IFU       &    56000\footnote{The GHOST radial velocity precision was measured from the obtained sky data to be equal to 0.04 {\kms}.}  &                  03/07/24       & 3$\times$508               & $\sim$0.7  \\
gb        & 17:24:35 & -39:00:26 & GHOST/Gemini & double IFU       &          56000 &          01/06/24       & 3$\times$508               &  $\sim$0.8 \\
gc        & 17:24:31 & -38:59:56 & GHOST/Gemini & double IFU        &      56000 &              03/07/24       & 3$\times$508               & $\sim$0.7  \\
gd        & 17:24:29 & -38:59:44 & GHOST/Gemini & double IFU        &      56000 &              02/07/24       & 3$\times$508               & $\sim$0.7  \\
ge        & 17:24:26 & -38:59:09 & GHOST/Gemini & double IFU        &         56000 &           01/06/24       & 3$\times$508               & $\sim$0.8 \\
gf        & 17:24:25 & -38:59:31 & GHOST/Gemini & double IFU        &       56000 &             03/07/24       & 3$\times$508               & $\sim$0.7  \\
gc\_sky   & 17:24:22 & -38:58:55 & GHOST/Gemini & double IFU       &      56000 &              03/07/24       & 3$\times$508               & $\sim$0.7 \\
gd\_sky   & 17:24:42 & -39:00:11 & GHOST/Gemini & double IFU        &     56000 &               02/07/24      & 3$\times$508              & $\sim$0.7  \\
fors2blue & 17:24:29 & -39:00:07 & FORS2/VLT    & spec: E2V, 1200B+197 &  1420        & 31/05/19        & 3$\times$900               & $\sim$0.7  \\
fors2red  & 17:24:29 & -39:00:10 & FORS2/VLT    & spec: MIT, 600RI+19   &  1000      & 02/06/19       & 3$\times$600               & $\sim$1    \\
fors2u    & 17:24:29 & -39:00:07 & FORS2/VLT    & imag: E2V, u-high     &       & 31/05/19       & 3, 40               & $\sim$0.7 \\
fors2b    & 17:24:29 & -39:00:07 & FORS2/VLT    & imag: E2V, b-high  &         & 31/05/19       & 3, 15               & $\sim$0.7  \\
fors2v    & 17:24:29 & -39:00:07 & FORS2/VLT    & imag: E2V, v-high     &        & 31/05/19       & 3, 40               & $\sim$0.7  \\
fors2R    & 17:24:29 & -39:00:10 & FORS2/VLT    & imag: MIT, Bessel R sp. & & 02/06/19       & 5, 3$\times$120            & $\sim$1    \\
fors2I    & 17:24:29 & -39:00:10 & FORS2/VLT    & imag: MIT, Bessel I         & & 02/06/19       & 10, 5$\times$120           & $\sim$1   \\
\hline
\end{tabular}
\end{table*}

\section{Results} \label{sec:res}

In this Section we present our results concerning both the PN PHR~J1724-3859 and the cluster Trumpler~25. Our findings regarding the cluster Trumpler~25 are presented in Subsection~\ref{subsec:T25}, while in Subsections~\ref{subsec:PHR} and~\ref{subsec:CSPN} we report our results about the PN PHR~J1724-3859 and its CSPN, respectively.

\subsection{Properties of open cluster Trumpler~25} \label{subsec:T25}

Gaia DR3 \citep{2022A&A...667A.148G} provides accurate kinematic data for a very large sample of sources 
and has been used for the identification of open cluster members \citep{2023A&A...675A..68V}. 
There are 550 stars assigned 100~\% probability of being members of Trumpler~25 
\citep{2023A&A...675A..68V}. We use these for determining the cluster mean physical 
parameters. We find a mean pmRA= 0.315 $\pm$ 0.002 mas/yr, a mean pmDEC= -2.138 $\pm$ 0.002 mas/yr and 
a mean parallax of 0.414 $\pm$ 0.002 mas, which implies a cluster distance of 2417 $\pm$ 14 pc. 
Gaia DR3 radial velocities are available for 40 of these stars and from these we estimate a cluster weighted mean 
radial velocity of -25.5 {\kms} with a  weight standard deviation of 9.1 {\kms} and a measurement uncertainty of 0.1 {\kms}. 
We use these cluster mean values for the remainder of our study. This observed dispersion, when OCs are usually
expected to have an inherent velocity dispersion of $\sim1-3$~{\kms} is unusually high for an OC and may 
reflect a large binary fraction \citsee{2010AJ....139.1383G}, explaining the 5~{\kms} offset 
between the PN systemic velocity and the cluster average (see Section \ref{subsec:PHR}).

\begin{figure*}
\plotone{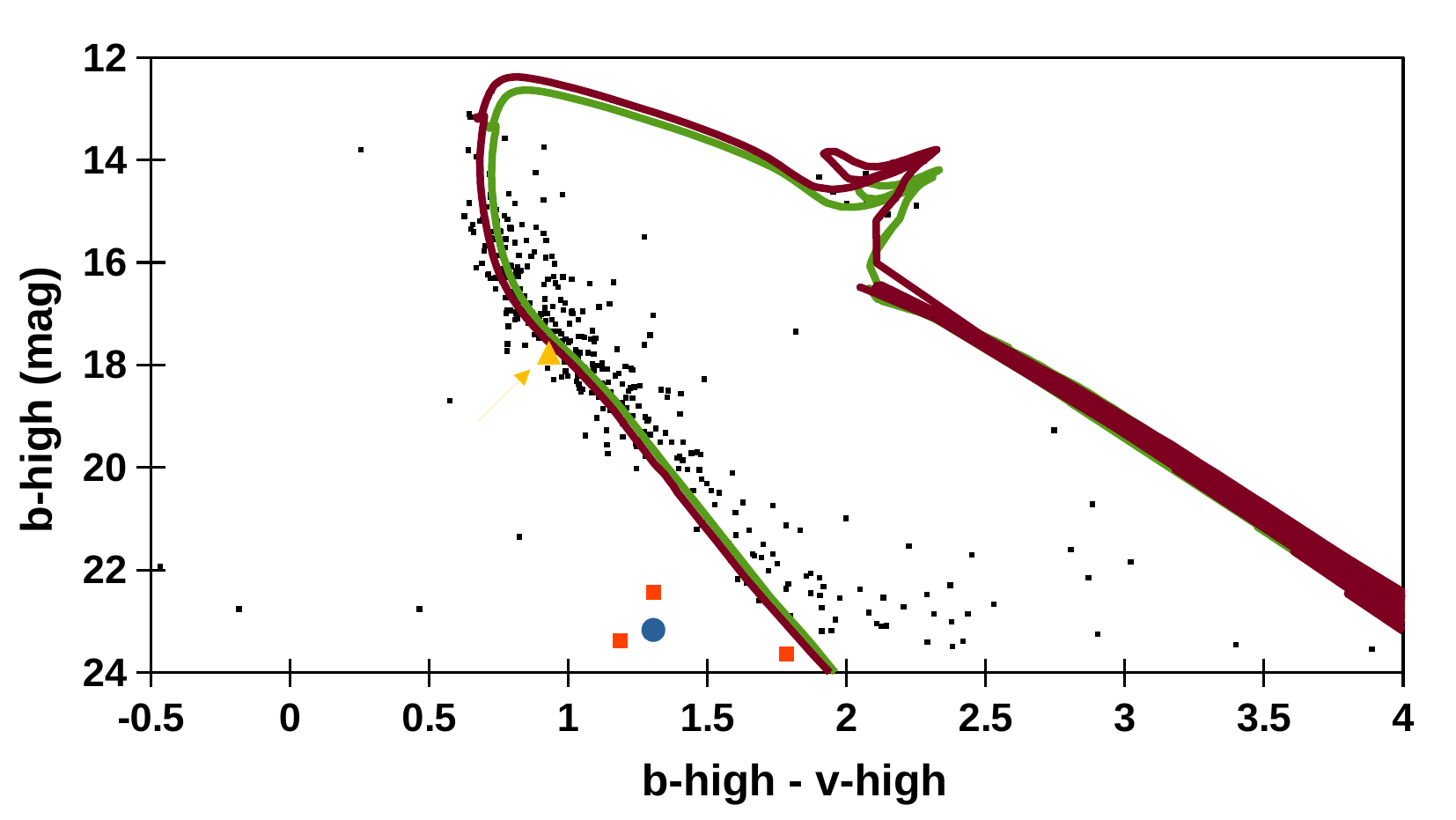}
\caption{The Trumpler~25 CMD constructed from FORS2 photometric b-high and v-high data of stars with cluster membership probability of 100~$\%$ \citep{2023A&A...675A..68V}. The green and brown lines are the theoretical isochrones produced from the cluster literature mean parameters and the ASteCa code, respectively. Although they both present a good fit, the ASteCa code presents a slightly better fit to our data. The data are not extinction corrected, with the interstellar extinction being incorporated in the fitted isochrones. The light red squares and blue circle depict the only blue stars within the nebula field. These were not part of the original cluster CMD that was used for the isochrone fit since they are too faint to be detected with Gaia. Their colour uncertainties (ranging from 0.15 to 0.22) do not affect their identification as evolved stars. The blue circle shows the location on the CMD of the blue star that is closest to the nebula apparent centre that we identify as the true CSPN. The yellow triangle (whose position in the plot is shown by the yellow arrow) indicates the location of the star that has been previously falsely identified as the CSPN by \citet{2021A&A...656A..51G}. It is evident that this star is not lying bluewards the MS (even if considering its colour uncertainty of 0.02) and towards fainter magnitudes as it would be expected for an evolved star as a CSPN.} 

\label{Fig3}
\end{figure*}

Our deep FORS2 photometry for stars with cluster membership probability of 100~\% \citep{2023A&A...675A..68V} 
has been employed for the construction of the cluster CMD (see Figure~\ref{Fig3}) and the determination of 
the cluster physical parameters.  Bessel R `special' and Bessel I imaging data cannot be optimally paired 
with the u-high, b-high and v-high data due to the different detectors and collimators used plus the 
different nights/seeing of their observations. As mentioned, u-high data were calibrated using literature 
photometry thus, they also cannot be optimally paired with data from other filters. For the aforementioned 
reasons, we use the more consistent b-high and v-high data for both the construction of the cluster CMD 
and the identification of the CSPN.

Padova theoretical isochrones \citep{2012MNRAS.427..127B} were fitted to our cluster CMD 
to determine the cluster physical parameters. Keeping the precise Gaia cluster distance determined above fixed, 
we employed the ASteCA code \citep{2015A&A...576A...6P} for determining the best fit isochrone and the 
remaining cluster physical parameters. We find that the cluster parameters for the isochrone 
\citep{2012MNRAS.427..127B,2014MNRAS.444.2525C,2014MNRAS.445.4287T,2015MNRAS.452.1068C,2017ApJ...835...77M,
2019MNRAS.485.5666P,2020MNRAS.498.3283P} that best fits our data are a cluster age of 112 $\pm$ 9 Myrs, 
a metal content of [Fe/H]= 0.17 $\pm$ 0.02 (implying a metallicity of Z= 0.023 $\pm$ 0.001 for $Z_\sun$=0.0152 
following \citet{2012MNRAS.427..127B}), a reddening of E[B-V]= 0.85 $\pm$ 0.01 (implying an extinction 
of $A_v$= 2.64 $\pm$ 0.04 using the \citet{1989ApJ...345..245C} extinction law with $Rv=3.1$) and a 
differential reddening of $DR = 0.21 \pm 0.02$. In Figure~\ref{Fig3} we show how the theoretical 
isochrones produced from using the previously calculated  mean literature 
values (see Section~\ref{sec:intro}, green isochrone) and the ASteCa code (brown isochrone) 
fit our FORS2 data for stars with cluster membership probability of 100~\%. The two isochrones 
are very similar, but the one calculated with the ASteCa code presents a slightly better fit. 

The ASteCa best fitted isochrone presents a turn-off (TO) point at B-V$=$0.685 and B$=$13.15 and, crucially, 
a TO mass of 4.88~$M_\sun$. Taking into consideration the time needed to leave the MS and pass 
through the Asymptotic Giant Branch (AGB) phase leads to a CSPN initial mass estimate of 
$ \rm M_{ini} = 5.12_{-0.15}^{+0.16}$. The initial mass errors reflect the errors in the cluster physical 
parameters. This is one of the highest CSPM progenitor masses determined to date with decent accuracy \citeg{2019NatAs...3..851F}.

\subsection{The Planetary Nebula PHR~J1724-3859}\label{subsec:PHR}
Due to cluster velocity dispersions being typically very low ($\leq$ 1-3\kms) 
with some exceptions \citep{2018A&A...619A.155S, 2021A&A...647A..19T}, a key property for the 
physical association of a PN with a cluster is the close agreement of their radial velocities. 
For this reason, we measured a precise nebular radial velocity using our high-resolution HRS and GHOST data. 
The resolution of GHOST was empirically estimated from the $\lambda6300\AA$ 
and $\lambda6364\AA$ sky lines of the gc\_sky and gd\_sky sky pointings, which are free from 
nebular emission. These should normally give a radial velocity equal to zero with deviations 
from this value due to the instrumental resolution. We derive a mean GHOST resolution and 
thus, radial velocity precision, of 0.04 {\kms}, more than enough for our purposes.  

For each one of our nebular pointings, we applied Gaussian fits to the stronger nebular emission 
lines (the [\ion{N}{ii}]$\lambda6548,6584$\AA~doublet and the H$\alpha$ line) using the IRAF 
task {\sc splot}. After determining their peak wavelengths and thus, their Doppler-shifts 
we determined the nebular radial velocity for each pointing. 

\begin{figure*}[ht!]
\centering

\gridline{
  \fig{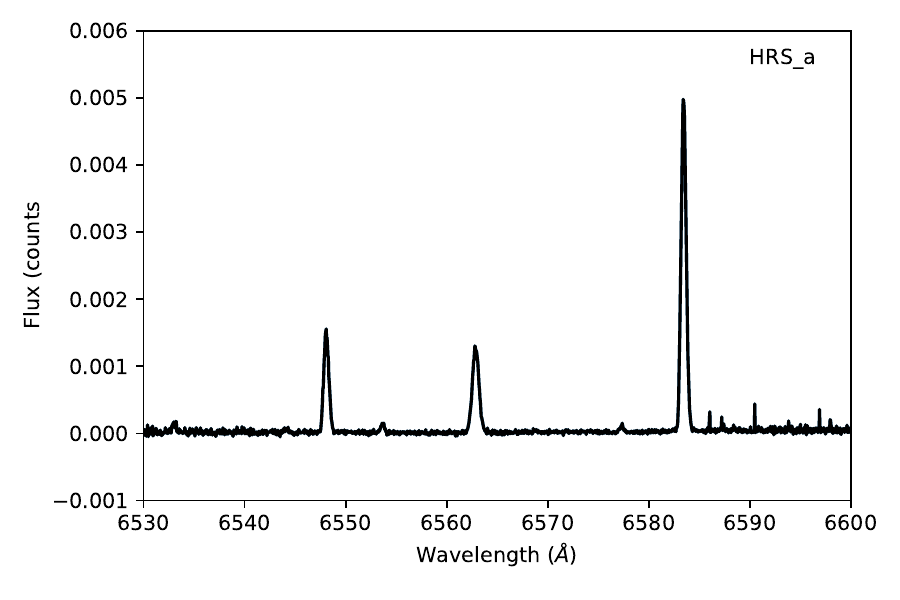}{0.32\textwidth}{pointing $ha$}
  \fig{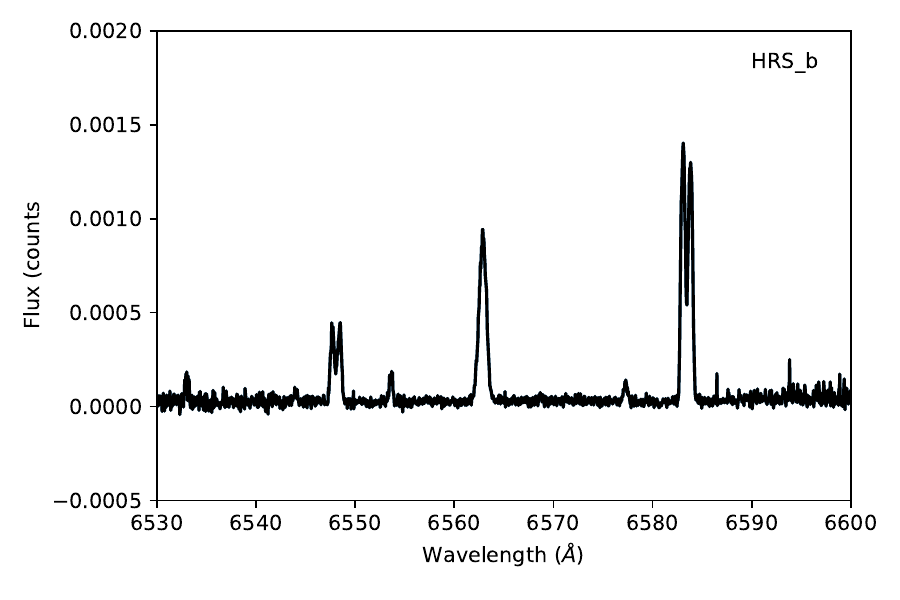}{0.32\textwidth}{pointing $hb$}
  \fig{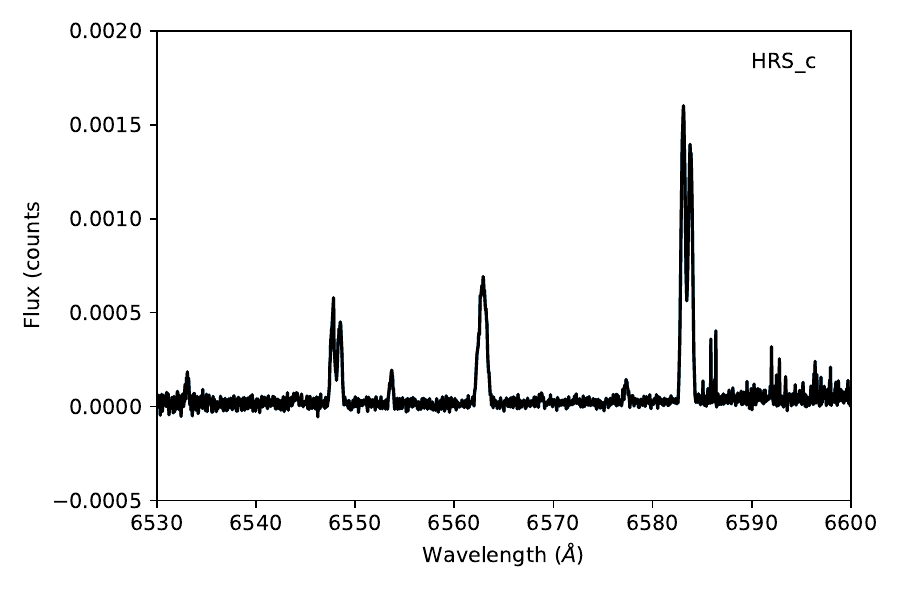}{0.32\textwidth}{pointing $hc$}
}

\gridline{
  \fig{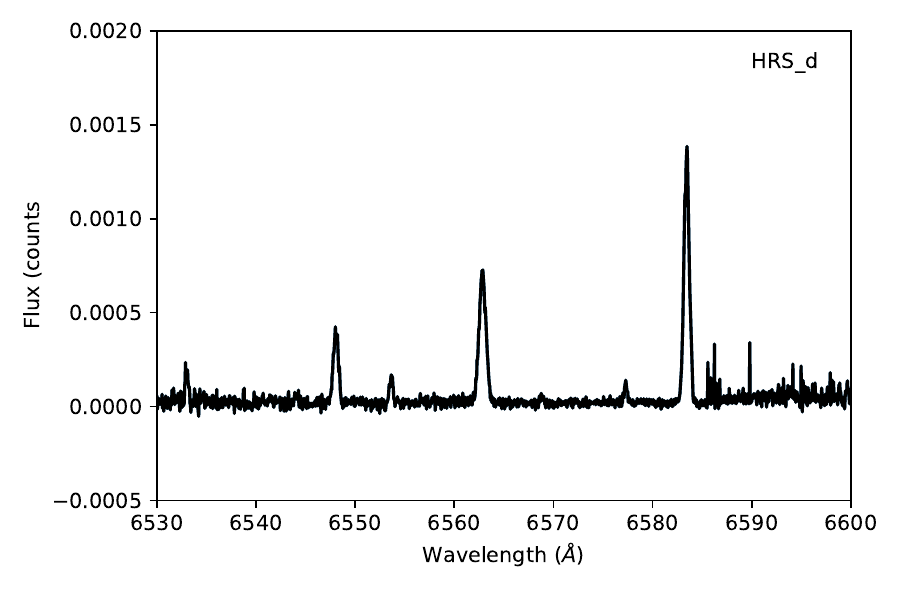}{0.32\textwidth}{pointing $hd$}
  \fig{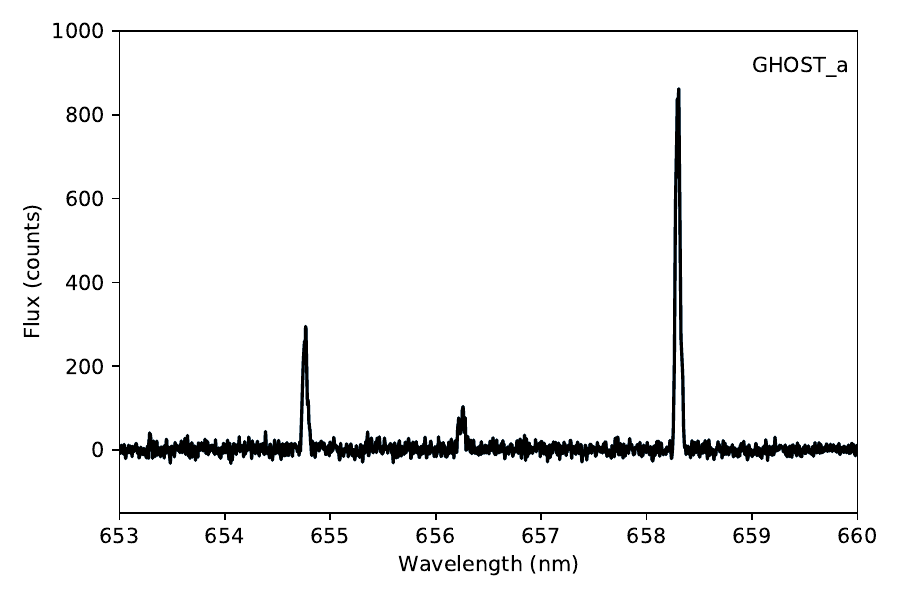}{0.32\textwidth}{pointing $ga$}
  \fig{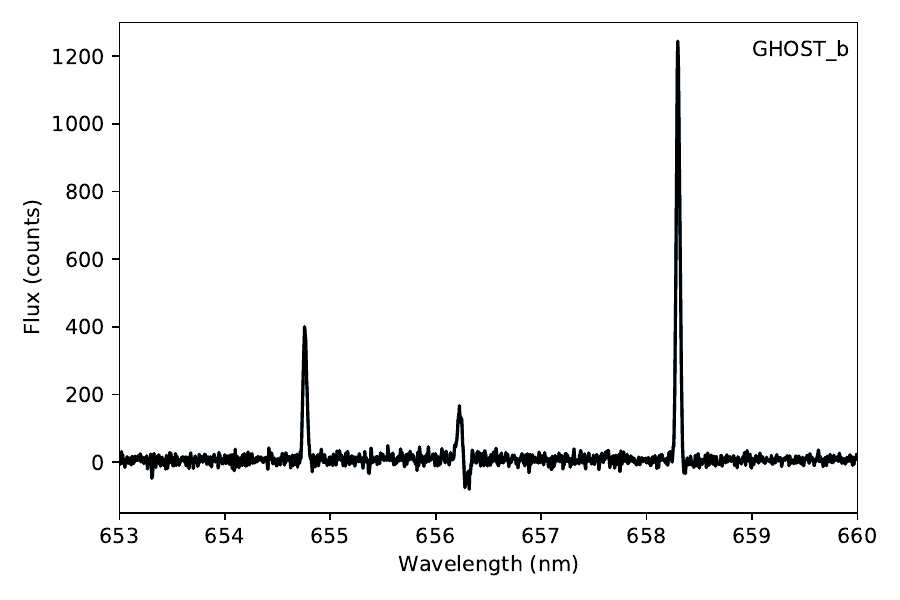}{0.32\textwidth}{pointing $gb$}
}

\gridline{
  \fig{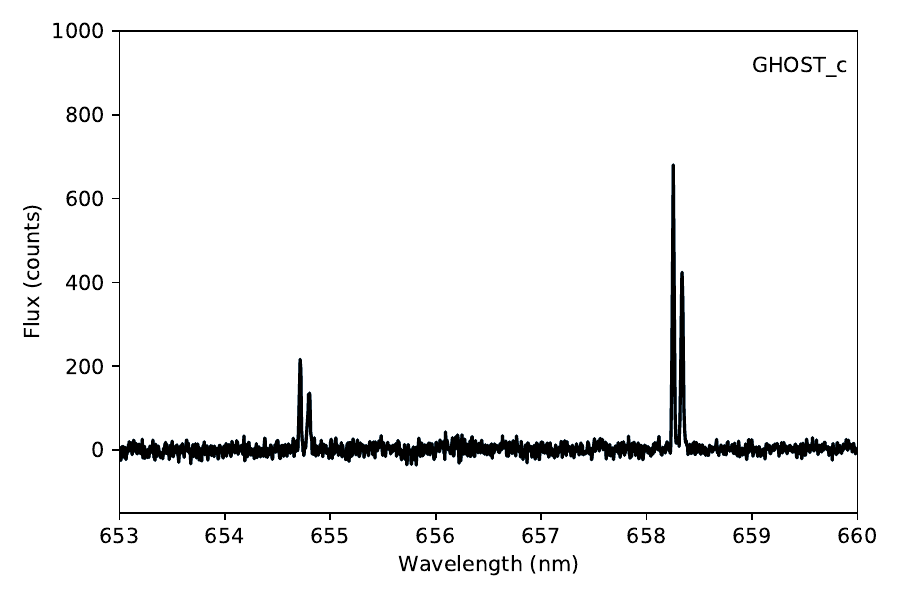}{0.32\textwidth}{pointing $gc$}
  \fig{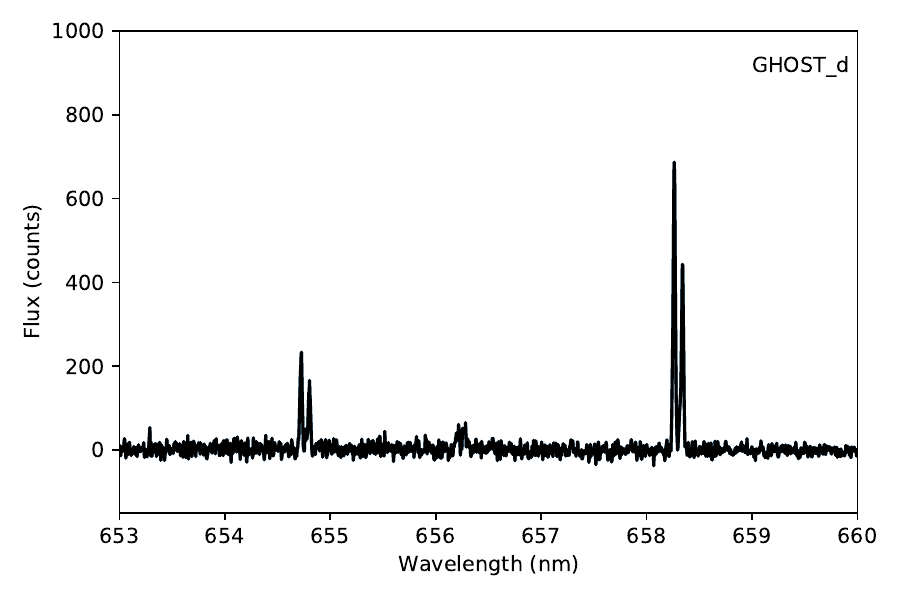}{0.32\textwidth}{pointing $gd$}
  \fig{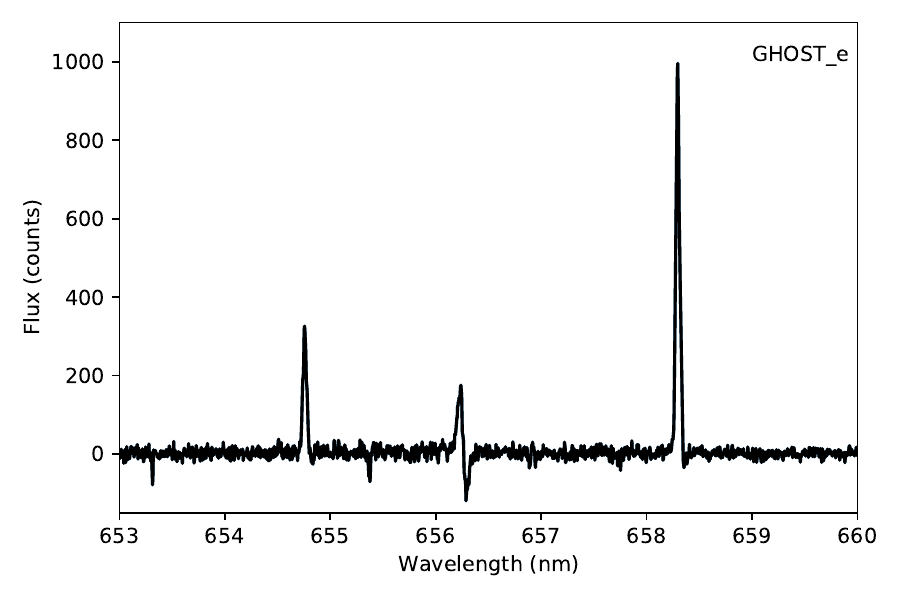}{0.32\textwidth}{pointing $ge$}
}

\gridline{
  \fig{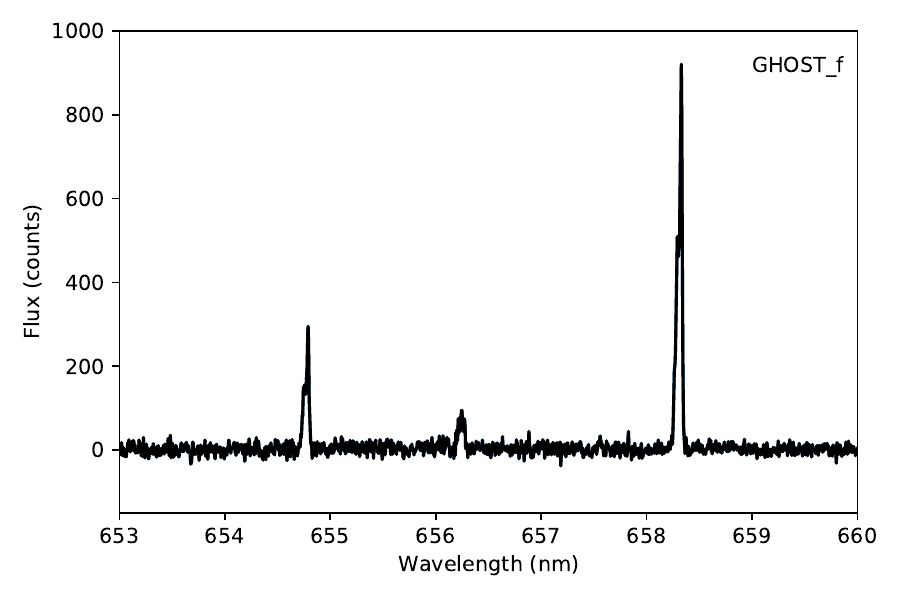}{0.32\textwidth}{pointing $gf$}
}

\caption{The HRS and GHOST reduced 1-D spectral data showing the [\ion{N}{ii}]$\lambda6548,6584\AA$ doublet and H$\alpha$ line for HRS pointings ha-hd and GHOST pointings ga-gf. The [\ion{N}{ii}]$\lambda6548,6584\AA$ lines present a prominent split in the spectra of pointings hb, hc, gc, gd and gf, while the same is true for the H$\alpha$ line in the spectrum of pointing gd.}
\label{Fig4}
\end{figure*}

The [\ion{N}{ii}]$\lambda6548,6584$\AA~and H$\alpha$ lines of pointing gd are split, while the same 
applies for just the [\ion{N}{ii}]$\lambda6548,6584$\AA~lines of pointings  hb, hc, gc and gf 
(see Figure~\ref{Fig4}). This simply reflects the internal nebular expansion velocity. 
Lines that are split into two components, often evident in high-resolution PNe spectra, allow not 
just an estimate of the PNe expansion velocity, but a more precise determination of radial velocity. 
This is because their peak is not biased towards the brightest component of front and back expansion
\citep{2012ApJ...751..116V}. The radial velocity for each pointing was estimated as the mean of 
the radial velocities calculated from each [\ion{N}{ii}]$\lambda6548,6584$\AA~and H$\alpha$ line present, while its assigned uncertainty implies the spread of the radial velocities calculated 
from the different emission lines. In the case of pointings gb and ge sky subtraction is problematic 
due to the sky exposures being acquired more than a month later and thus, for these pointings 
we excluded the H$\alpha$ line in the calculation of their radial velocities. In the case of 
split lines, we initially measured the peak of each component separately and then calculated 
their average. The H$\alpha$ line is not detected in pointing gc due to the relatively 
low surface brightness of the PN at this pointing. Table~\ref{Tab2} shows our radial velocity 
estimates for each pointing. The radial velocity spread in different pointings reflects 
the internal movement of the nebular gas.

%\begin{figure*}[!]
%\subfloat[pointing $ha$]{\includegraphics[width = 2.5in]{19_1_hrs.pdf}}
%\subfloat[pointing $hb$]{\includegraphics[width = 2.5in]{19_2_hrs.pdf}}
%\subfloat[pointing $hc$]{\includegraphics[width = 2.5in]{20_1_hrs.pdf}}\\
%\subfloat[pointing $hd$]{\includegraphics[width = 2.5in]{20_2_hrs.pdf}} 
%\subfloat[pointing $ga$]{\includegraphics[width = 2.5in]{ghost_a.pdf}} 
%\subfloat[pointing $gb$]{\includegraphics[width = 2.5in]{ghost_b.pdf}}\\
%\subfloat[pointing $gc$]{\includegraphics[width = 2.5in]{ghost_c.pdf}}
%\subfloat[pointing $gd$]{\includegraphics[width = 2.5in]{ghost_d.pdf}} 
%\subfloat[pointing $ge$]{\includegraphics[width = 2.5in]{ghost_e.pdf}} \\
%\centering
%\hspace{1in}\subfloat[pointing $gf$]{\includegraphics[width = 2.5in]{ghost_f.pdf}} 
%%\hspace{1.2in}
%\caption{The HRS and GHOST reduced 1-D spectral data showing the [\ion{N}{ii}]$\lambda6548,6584\AA$ 
%doublet and H$\alpha$ line for HRS pointings ha-hd and GHOST pointings ga-gf. The 
%[\ion{N}{ii}]$\lambda6548,6584\AA$ lines present a prominent split in the spectra of
%pointings hb, hc, gc, gd and gf, while the same is true for the H$\alpha$ line in the 
%spectrum of pointing gd.} 
%\label{Fig4}
%\end{figure*}

\begin{table*}[]
\centering
\caption{The derived PN radial and expansion velocities, [\ion{N}{ii}]/H$\alpha$ and 
[\ion{S}{ii}] $I(6731)/I(6716)$ ratios, and $N_e$ for each pointing }
\label{Tab2}
\begin{tabular}{cccccc}
\hline \hline
ID  & $v_{\rm r}$ $\pm$ $\sigma$ ({\kms}) & $v_{\rm exp}$ ({\kms}) & {[}NII{]}/H$\alpha$\footnote{Uncorrected for 
extinction. Corresponding ratios not estimated for: a) pointings gb and ge due to the problematic sky subtraction 
of their frames, b) pointing gc due to the absence of the $H\alpha$ line, c) all HRS pointings since HRS spectra 
is neither sky subtracted not flux calibrated.} & $I(6731)/I(6716)$\footnote{HRS pointings have no assigned 
$I(6731)/I(6716)$ ratios and $\rm N_e$ due to the corresponding data being neither sky subtracted nor flux 
calibrated, rendering the calculation of these parameters not optimal.} & $\rm N_e$\footnote{Assuming $T_e$= 
10000 K} ($\rm cm^{-3}$) \\
\hline 
ha    & -19.6$\pm$1.9                                        & 28                              &                     &                   &                           \\
hb    & -20.1$\pm$2.2                                        & 17                              &                     &                   &                           \\
hc    & -19.6$\pm$1.9                                        & 17                              &                     &                   &                           \\
hd    & -20.3$\pm$1.6                                        & 28                              &                     &                   &                           \\
ga    & -19.0$\pm$4.7                                          & 29                              & 11.6                & 0.71$\pm$0.09     & \textless 100             \\
gb    & -21.1$\pm$1.3                                        & 16                              &                     & 0.61$\pm$0.07     & \textless 100             \\
gc    & -22.8$\pm$1.4                                        & 19                              &                     & 0.98$\pm$0.11     & 657$\pm$330               \\
gd    & -19.5$\pm$1.3                                        & 17                              & 9.6                 & 0.53$\pm$0.13     & \textless 100             \\
ge    & -22.3$\pm$1.5                                        & 18                              &                     & 0.65$\pm$0.08     & \textless 100             \\
gf    & -15.8$\pm$1.7                                        & 19                              & 9.3                 & 0.87$\pm$0.12     & 378$\pm$270               \\
fors2 &                                                      &                                 & 12.6\footnote{Over-subtraction of the H$\alpha$ line due to sky contamination. The true [NII]/H$\alpha$ ratio is likely lower than this.}                & 0.75$\pm$0.16     & \textless 100         \\   
\hline
\end{tabular}
\end{table*}

The systemic nebular radial velocity, estimated as the mean from all nebular pointings is $-20.0\pm0.7$ {\kms} 
with a standard deviation of 1.9 {\kms}. Taking into consideration only the emission lines that present a split, 
which have more precise peaks (see above), we estimate a systemic nebular radial velocity of $-19.4\pm0.7$ {\kms} 
very close to the previous value. Taking into account only the HRS and GHOST data separately, we estimate a PN 
radial velocity of $-19.9\pm1.0$ {\kms} and $-20.1\pm1.0$ {\kms} respectively, confirming the accuracy of our 
method and so the robustness of the PN radial velocity estimate.

We also measured the nebular expansion velocity for each HRS and GHOST pointing (see Table~\ref{Tab1}) from the 
split of the [\ion{N}{ii}] lines and the full width half maximum (FWHM) of the H$\alpha$ line, taking into 
consideration the thermal broadening estimated using the $\lambda6300\AA$ night-sky line. To specify, the 
expansion velocity for pointings hb, hc, gc and gd (which present a split and have problematic or no H$\alpha$ 
detection) was measured from their [\ion{N}{ii}] line splitting and for pointigs ha, hd, ga, gb, ge (that do not 
present a split) from the FWHM of their H$\alpha$ line, while for pointing gf (which presents both a prominent 
split of the [\ion{N}{ii}] lines and a solid H$\alpha$ detection) as the average calculated with both methods. 
HRS and GHOST can attain a similar velocity precision and thus, using the data from all HRS and GHOST pointings, 
we calculate a mean nebular expansion velocity of 21~{\kms} with 
$\sigma= 5${\kms}. Applying the correction factor from \citet{2013A&A...558A..78J} to account for the fact that 
the velocity of the nebular outer edges cannot be determined from spectral data, we estimate a corrected nebular 
expansion velocity of 31~{\kms} with $\sigma= 8$~{\kms}, completely typical for a PN.  
 
\begin{figure*}[ht!]
\centering

\gridline{
  \fig{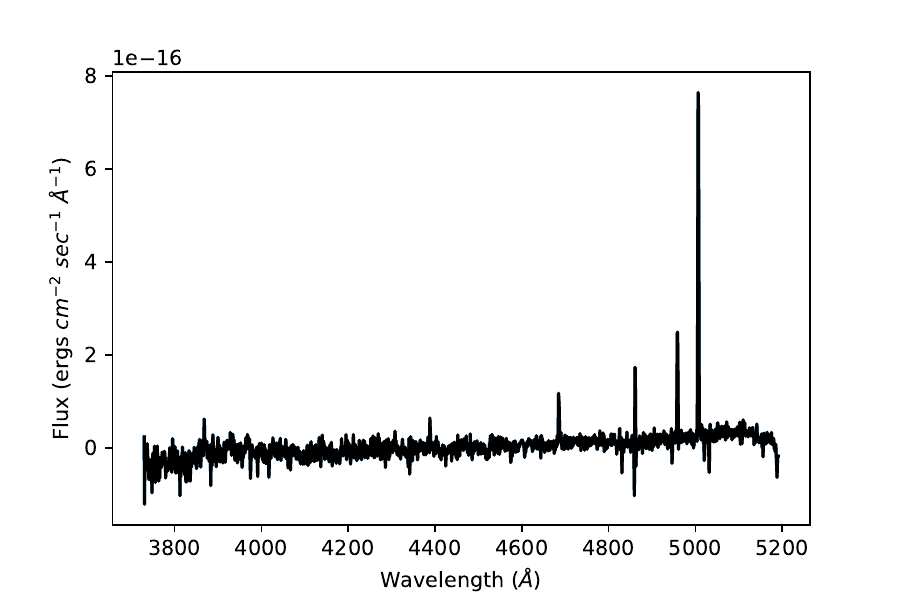}{0.45\textwidth}{Blue arm}
  \fig{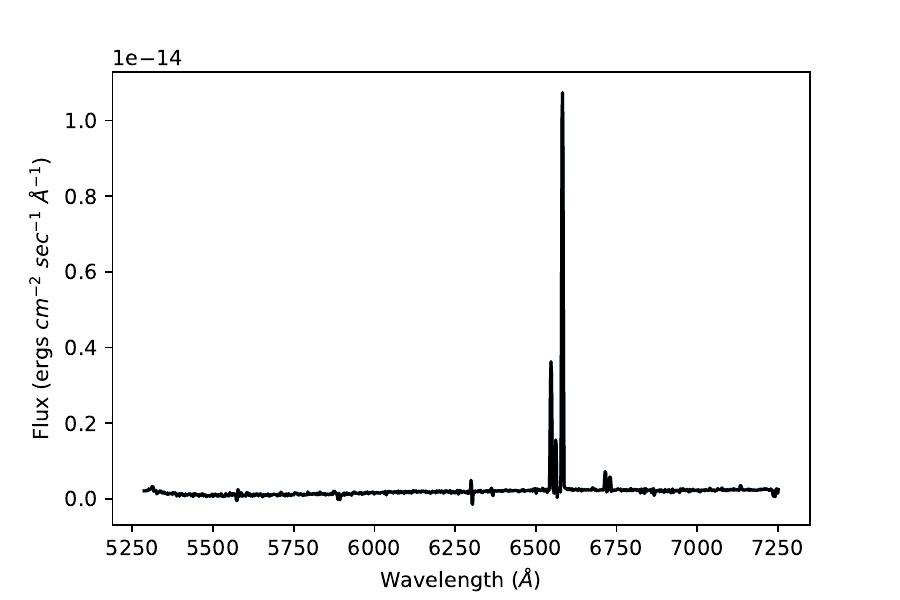}{0.45\textwidth}{Red arm}
}

\caption{The FORS2 blue arm (left panel) and red arm (right panel) 1-D spectra. Modest sky line over-subtraction 
is evident, especially in the blue arm. The high excitation He~II $\lambda4686\AA$ emission line and $H\beta$, 
[\ion{O}{iii}]$\lambda4959,5007\AA$, [\ion{N}{ii}] $\lambda6548,6563\AA$, $H\alpha$ and [\ion{S}{ii}] 
$\lambda6716,6731\AA$ lines are clearly identified.}
\label{Fig5}
\end{figure*}

Due to the problematic sky subtraction (see Section~\ref{sec:obs}) of our FORS2 spectra (see Figure~\ref{Fig5}), 
we cannot use the FORS2 data for estimating the PN reddening. HRS and GHOST spectra are not flux 
calibrated, so are also inadequate for calculating the nebular reddening.
Our spectral data, used for the following calculations, are not corrected for interstellar extinction 
due to the lack of a reliable PN reddening measured directly from them. Since the spectral lines used for 
the calculation of the corresponding line ratios are sufficiently close, this is not expected to 
significantly affect our results.

%\begin{figure}
%\subfloat{\includegraphics[width = 3.6in]{FORS2_blue_spectra.pdf}}
%\subfloat{\includegraphics[width = 3.6in]{FORS2_red_spectra.pdf}} \\
%\caption{The FORS2 blue arm (left panel) and red arm (right panel) 1-D spectra. Modest sky line over-subtraction 
%is evident, especially in the blue arm. The high excitation He~II $\lambda4686\AA$ emission line and $H\beta$, 
%[\ion{O}{iii}]$\lambda4959,5007\AA$, [\ion{N}{ii}] $\lambda6548,6563\AA$, $H\alpha$ and [\ion{S}{ii}] 
%$\lambda6716,6731\AA$ lines are clearly identified.} 
%\label{Fig5}
%\end{figure}

\begin{figure*}[ht!]
\centering

\gridline{
  \fig{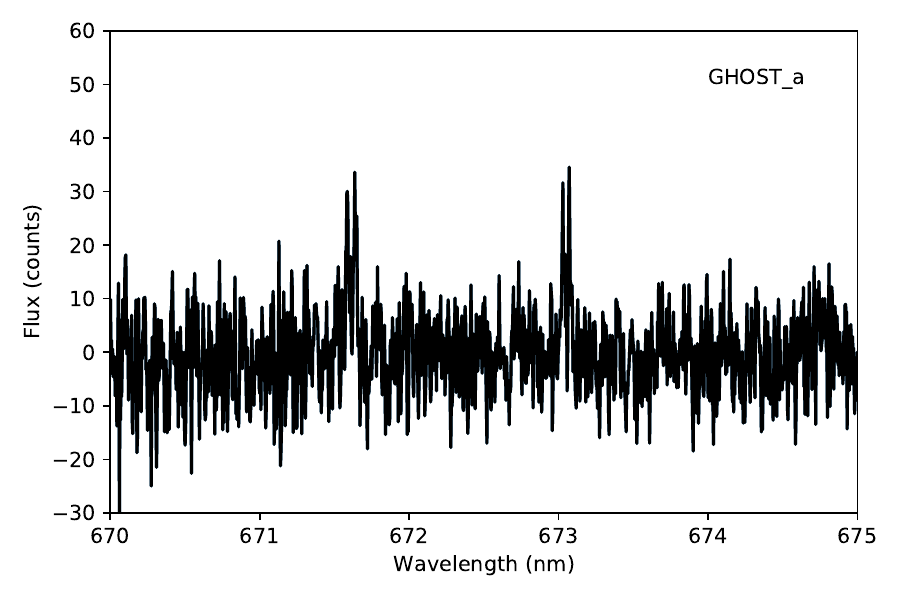}{0.30\textwidth}{pointing $ga$}
  \fig{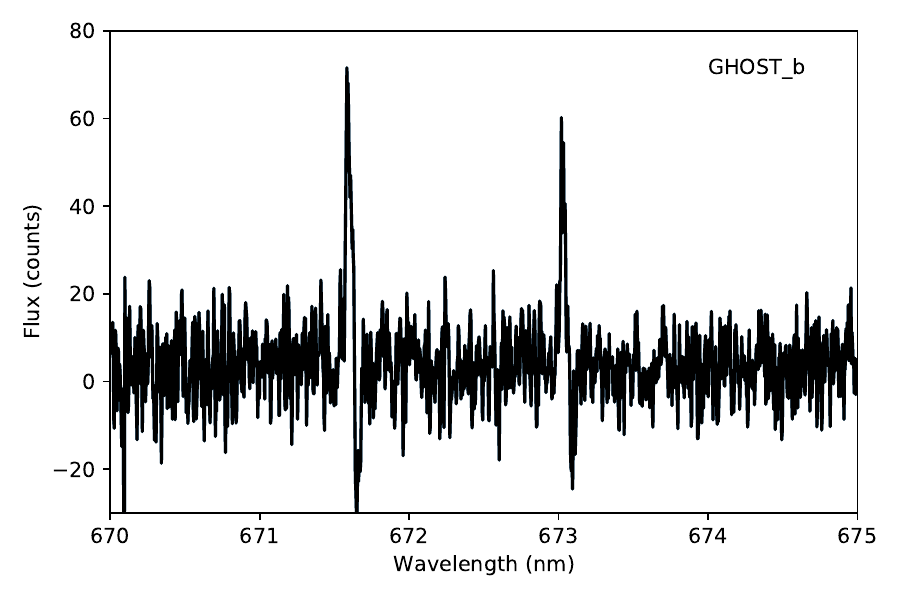}{0.30\textwidth}{pointing $gb$}
  \fig{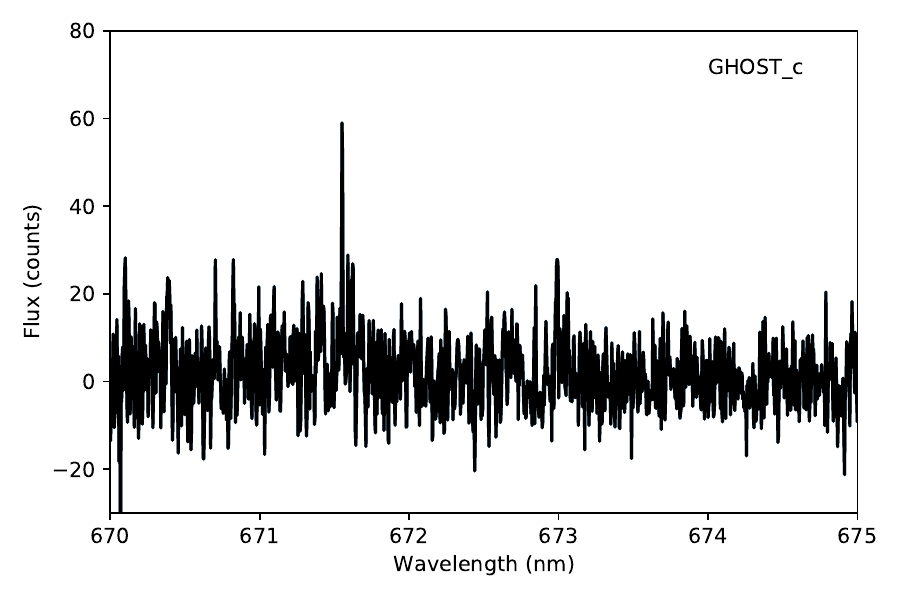}{0.30\textwidth}{pointing $gc$}
}

\gridline{
  \fig{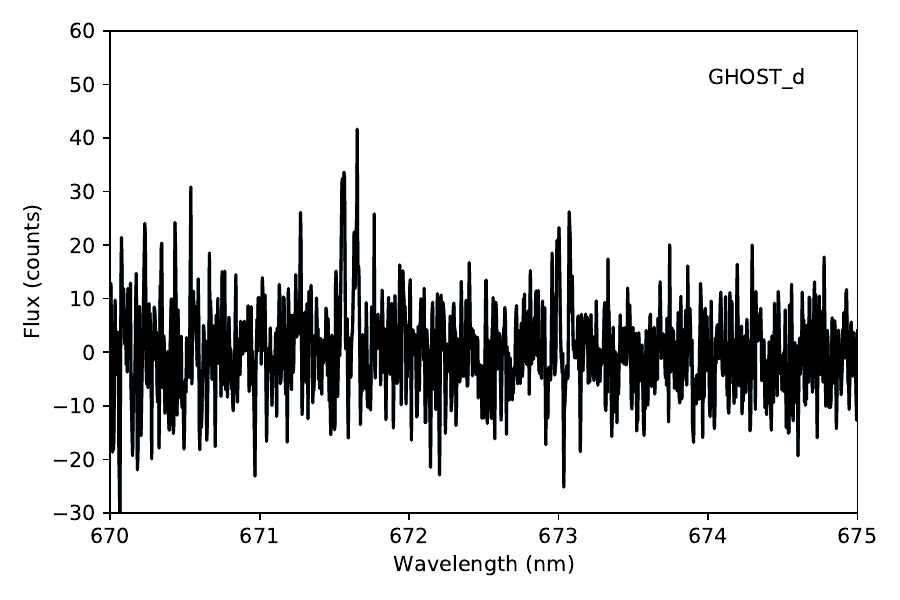}{0.30\textwidth}{pointing $gd$}
  \fig{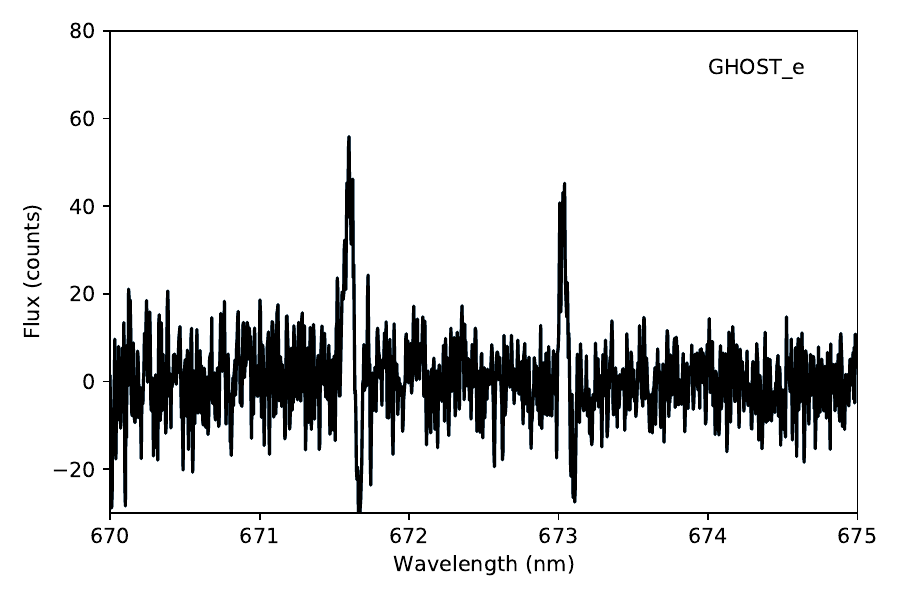}{0.30\textwidth}{pointing $ge$}
  \fig{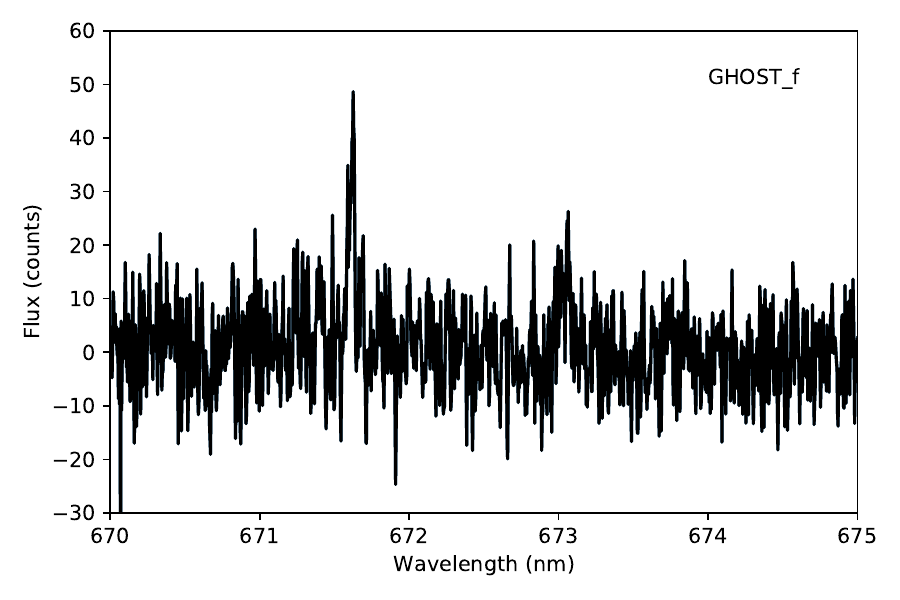}{0.30\textwidth}{pointing $gf$}
}

\caption{The 1-D spectra of all GHOST pointings, depicting the [\ion{S}{ii}] doublet. 
A small over-subtraction issue is evident in the spectra of pointings gb and ge.}
\label{Fig6}
\end{figure*}

For estimating the nebular electron density $N_e$, Gaussian fits were also applied to the [\ion{S}{ii}] 
$\lambda6716,6731\AA$ doublet that was detected in our GHOST (see Figure ~\ref{Fig6}) and FORS2 spectra, in 
order to measure the corresponding [\ion{S}{ii}]$\lambda6731/6716\AA$ ratios. %The HRS spectra is neither flux calibrated nor sky subtracted, but this should not affect the nebular [\ion{S}{ii}]$\lambda6731/6716\AA$ ratios given the lines are so close together in wavelength, though our other observations are better for this. 
HRS will not be used for $N_e$ because it is neither flux-calibrated nor sky-subtracted. While 
GHOST data are not flux calibrated and FORS2 data suffer from modest sky subtraction issues, these 
issues alone are not expected to substantially affect the estimation of the [\ion{S}{ii}]$\lambda6731/6716\AA$ 
ratio, since the [\ion{S}{ii}] lines are very close to each other and their ratio is not significantly affected 
by sky emission at different velocity. In the spectrum of pointing gd, both [\ion{S}{ii}] $\lambda6716\AA$ and 
$\lambda6731\AA$ lines present a split as for other lines and so the emission of each component was measured 
separately and then added to estimate their total emission. The nebular electron density was estimated for each 
GHOST pointing and FORS2 data from their [\ion{S}{ii}]$\lambda6731/6716\AA$ ratio and the {\sc temden} task in 
{\sc PyNeb} \citep{2015A&A...573A..42L}, assuming a $T_e$=10000 K (see Table~\ref{Tab2}). The lines used for the derivation of the electron density are not corrected for extinction. This does not limit the analysis as their wavelengths are sufficiently close that extinction should not effect the observed ratio. The [\ion{S}
{ii}]$\lambda6731/6716\AA$ doublet is not sensitive enough to derive a precise electron density close to the low 
density limit \citep{2006agna.book.....O}. For most of our pointings the PN nebular density was found to be at the low density limit of $\rm 
N_e < 100$ $\rm cm^{-3}$. Two of our pointings (gc and gf) do have a larger electron density due to local 
density variations. The nebular mean electron density was estimated using the mean [\ion{S}
{ii}]$\lambda6731/6716\AA$ ratio, calculated from all GHOST pointings plus FORS2. The mean [\ion{S}
{ii}]$\lambda6731/6716\AA$ ratio was found to be equal to 0.73 $\pm$ 0.04 and thus, the mean electron density is 
also estimated to be $\rm N_e < 100$ $\rm cm^{-3}$ and close to the low density limit.

%\begin{figure*}[!]
%\subfloat[pointing $ga$]{\includegraphics[width = 2.4in]{ghost_a_sii.pdf}} 
%\subfloat[pointing $gb$]{\includegraphics[width = 2.4in]{ghost_b_sii.pdf}}
%\subfloat[pointing $gc$]{\includegraphics[width = 2.4in]{ghost_c_sii.pdf}}\\
%\subfloat[pointing $gd$]{\includegraphics[width = 2.4in]{ghost_d_sii.pdf}} 
%\subfloat[pointing $ge$]{\includegraphics[width = 2.4in]{ghost_e_sii.pdf}} 
%\subfloat[pointing $gf$]{\includegraphics[width = 2.4in]{ghost_f_sii.pdf}} \\
%%\hspace{1.2in}
%\caption{The 1-D spectra of all GHOST pointings, depicting the [\ion{S}{ii}] doublet. 
%A small over-subtraction issue is evident in the spectra of pointings gb and ge. } 
%\label{Fig6}
%\end{figure*}

The [\ion{N}{ii}]/H$\alpha$ line ratio was also estimated from the spectral data of some of the 
GHOST pointings, and the FORS2 data (see Table~\ref{Tab2}). HRS pointings were not employed for 
this purpose due to their limitations in the calculation of line ratios (see above). Pointings gb 
and ge suffer from over-subtraction of the geocoronal $H\alpha$ emission line due to problematic sky subtraction in the case of these two science pointings, and there is no 
$H\alpha$ detection in the spectrum of pointing gc, so the [\ion{N}{ii}]/H$\alpha$ line ratio 
was also not derived for these. From the ga, gd and gf GHOST pointings we estimate a mean 
[\ion{N}{ii}]/H$\alpha$=10.2, while from our FORS2 data we estimate a [\ion{N}{ii}]/H$\alpha$=12.6. 
In the case of the FORS2 data, the quoted [\ion{N}{ii}]/H$\alpha$ ratio is likely higher than the 
true value as a result of the over-subtraction of the H$\alpha$ line in our data due to the sky 
contamination present in the FORS2 science frames. From our FORS2 data we, also, estimate a 
[\ion{O}{iii}]/H$\beta$ line ratio of 8, which also should be considered as an upper limit and 
note the presence of the high excitation He II $\lambda4686\AA$ emission line (see Figure ~\ref{Fig5}). 
For a further discussion about the nebular estimated emission line ratios, see Section~\ref{sec:analysis}.

\subsection{The Central Star of PHR~J1724-3859}\label{subsec:CSPN}
 \citet{2021A&A...656A..51G} and \citet{2021A&A...656A.110C} have reported the identification of many putative 
CSPN using Gaia data, including that for PHR~J1724-3859. The yellow triangle (marked by the yellow arrow) in 
Figure ~\ref{Fig3} indicates the location of the CSPN identified by \citet{2021A&A...656A..51G} on our FORS2 
CMD. It is noted that in our CMD (see Figure~\ref{Fig3}) the identified star does not lie where it should if being the true CSPN (i.e. in the fainter part of the diagram and bluewards from the MS, as expected for evolved stars). Although this star is the bluest Gaia detected star in the field, and indeed the major basis on which CSPN were chosen 
\citep{2021A&A...656A..51G}, the true CSPN is fainter. Indeed, \citep{2022Galax..10...32P} have shown that many 
true CSPNe are beyond GAIA limits (g$\sim$21). The approach adopted by both 
\citet{2021A&A...656A..51G} and  \citet{2021A&A...656A.110C} leads to an estimated 20$\%$ CSPN misidentification 
rate \citep{2022Galax..10...32P}. For these reasons, we make an effort to identify afresh the true CSPN of 
PHR~J1724-3859 using our much deeper FORS2 data. 

\begin{figure*}
\plotone{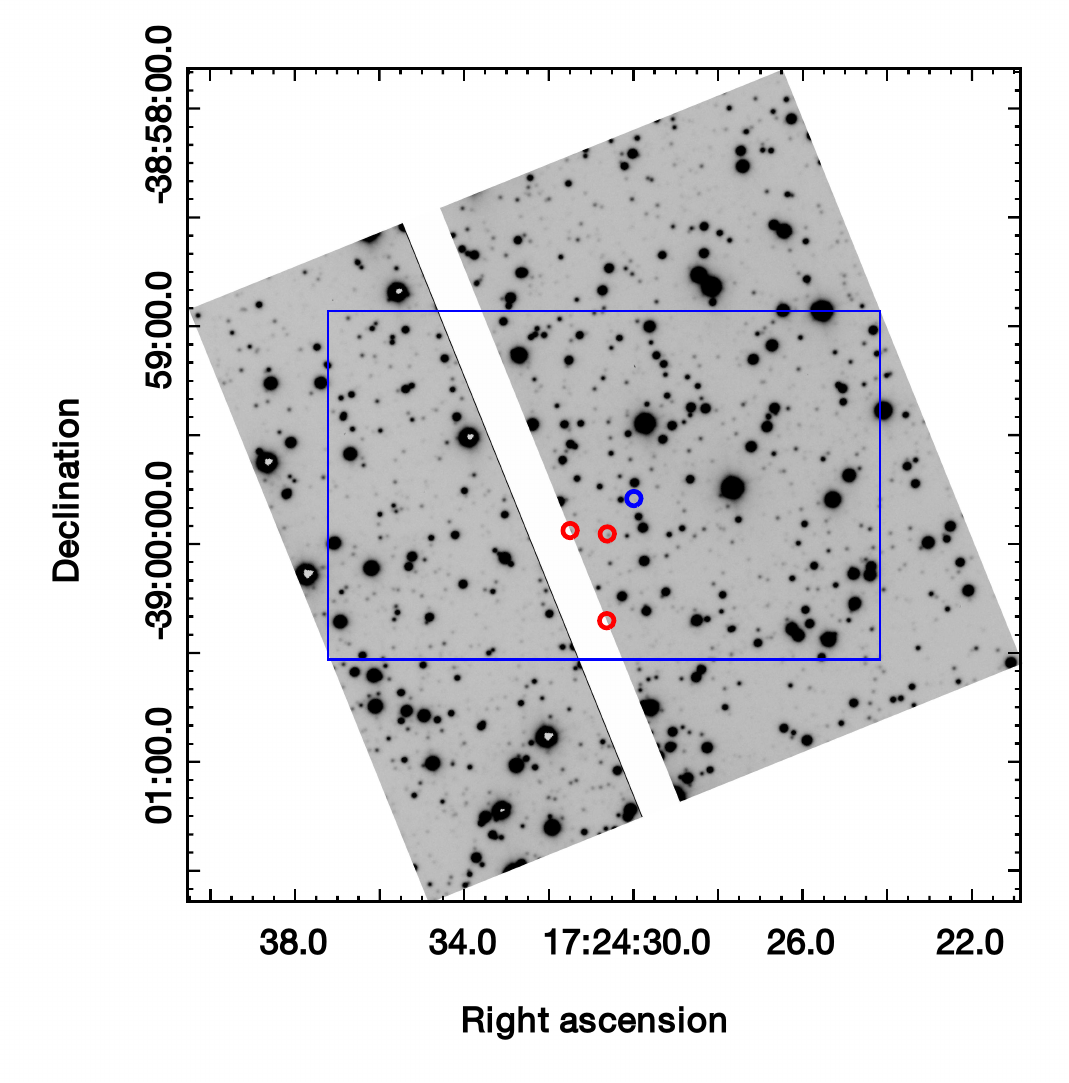}
\caption{A v-high 30$\times$30 $arcsec^2$ FORS2 image centred on the PN's apparent centre. The blue rectangle 
indicates the position and size of the PN. The blue and red circles show all blue stars in the nebular field. 
The blue circle indicates the blue star we have newly identified as the CSPN that lies closest to the nebular 
apparent centre.}
\label{Fig7}
\end{figure*}

\begin{figure}
\plotone{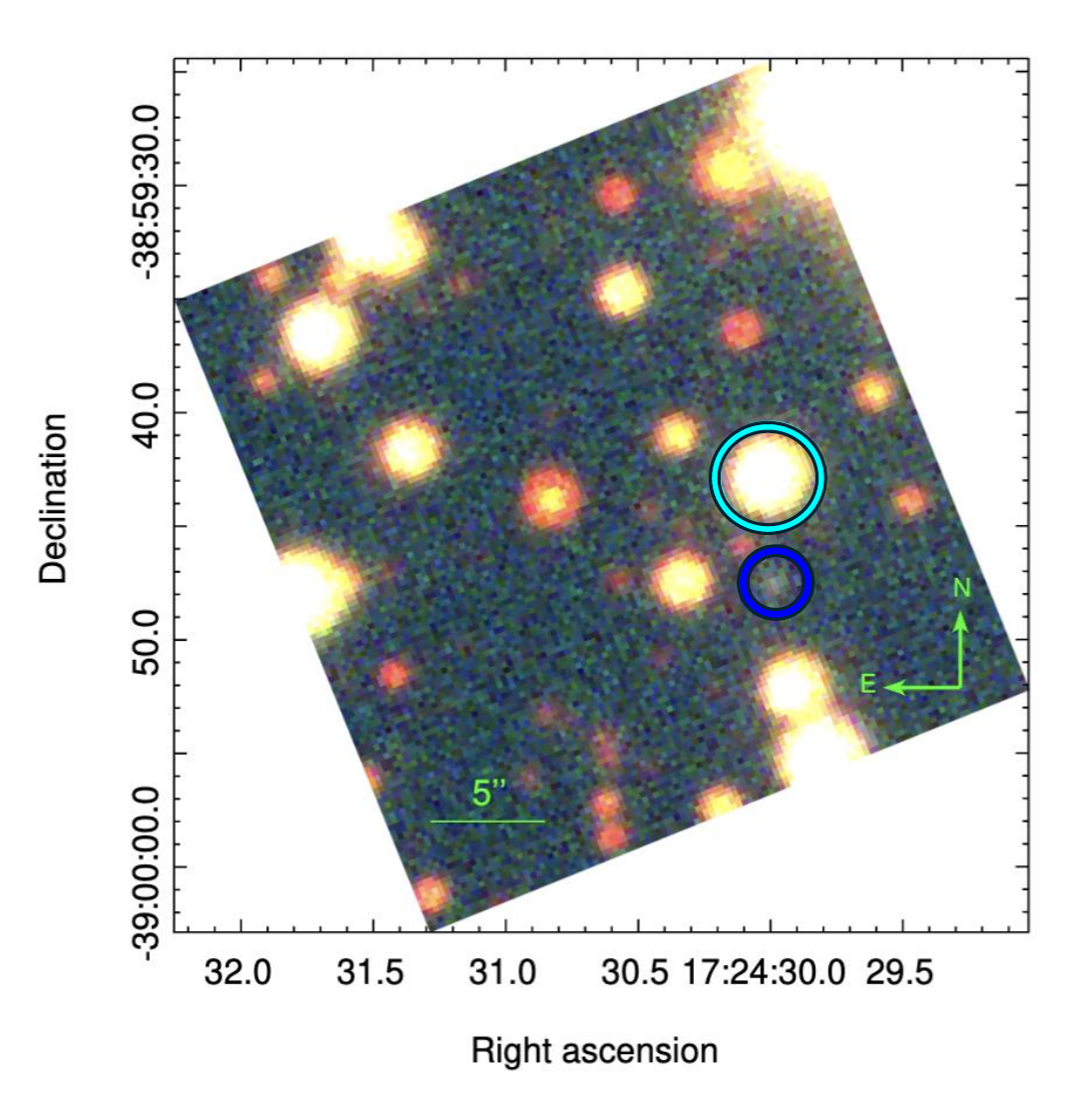}

\caption{A 30$\times$30 $arcsec^2$ FORS2 RGB (R is v-high, G is b-high and B is u-high) image showing the 
identified CSPN with J2000 position RA: 17h24m29.904s, DEC: -38\degree59'47.15".
The dark blue circle shows the location of the true CSPN newly identified here, while the light blue circle shows the 
one identified by \citet{2021A&A...656A..51G}. The two stars are within 2 arcsec of each other.}
\label{Fig8}
\end{figure}

By over-plotting all FORS2 photometrically identified stars in the nebula field on our cluster CMD, we note 
stars that lie bluer than the MS where their colours and magnitudes are consistent with these of 
white dwarfs (as indicated with a blue circle and light red squares in Figure ~\ref{Fig3}). We found only four 
blue stars in the nebular field (see Figure~\ref{Fig7}). The one that lies closest to the projected apparent PN 
geometric centre (indicated with a blue circle in Figure~\ref{Fig3} and Figure~\ref{Fig7}), is 
deemed the true CSPN with J2000 position RA: 17h24m29.904s, DEC: -38\degree59'47.15". This is only $\sim2.6$~arcseconds to the West from our estimated geometric centre for a PN 
that is 156$\times$96 arcseconds in angular extent. The previously identified CSPN is 
offset $\sim7$~arseconds from the PN centroid to the North West. The blue stars identified here are not 
originally part of the cluster CMD because they are beyond the reach of Gaia. The %much brighter 
brightest 
putative CSPN identified by \citet{2021A&A...656A..51G} is within 2 arcsec of our newly identified CSPN 
(see Figure~\ref{Fig8}). 

SExtractor can perform photometric measurements for many sources automatically and thus, 
is an efficient method for constructing cluster CMDs and to explore the deviation of stellar colours from the MS. However, the calculated magnitudes may not be precise for very dim stars in a crowded field, as is the case of 
the newly identified CSPN. Therefore, the b-high and v-high magnitudes of our CSPN, were measured again manually 
with aperture photometry and the IRAF task {\sc daophot/qphot}. This is after applying aperture correction and 
carefully selecting the appropriate aperture radius (1 arcsec), annulus (1 arcsec) and dannulus (0.5~arcsec) 
for the specific star and using the appropriate zeropoints for each filter. Using the same parameters and 
method, we also estimated the u-high, Bessel R special and Bessel I magnitudes. The estimated b-high, v-high, 
Bessel R special and Bessel I  magnitudes were then corrected for atmospheric extinction, while a colour term 
correction was also applied (see Section~\ref{sec:obs}). The measured u-high magnitude was calibrated by 
cross-correlation with the \citet{2010PASP..122.1437P} U magnitudes, as before (see Section~\ref{sec:obs}). 
We find u-high = 24.5 $\pm$ 0.4 mag, b-high = 24.5 $\pm$ 0.3 mag, v-high = 23.9 $\pm$ 0.1 mag, Bessel 
R special = 23.5 $\pm$ 0.4 mag and Bessel I = 22.8 $\pm$ 0.2 mag. The more precise manually measured b-high 
and v-high CSPN magnitudes are somewhat fainter than those measured with SExtractor due to the special care 
taken to avoid contamination from nearby stellar emission.
  
Theoreticaly, a CSPN effective temperature for a PN that has He II $\lambda4686\AA$ high-excitation line emission, like PHR~J1724-
3859, is expected to be between $\sim$50 to up to 300~kK \citep[see,][]{2016A&A...588A..25M, 2017ApJ...836...93J} though no 
known PN has a temperature beyond 250~kK. Under this consideration and using the Astrolib PySynphot 
package\footnote{Lim, P. L., Diaz, R. I., \& Laidler, V. 2015, PySynphot User’s Guide (Baltimore, MD: STScI), {\url https://pysynphot.readthedocs.io/en/latest/} } 
\citep{2013ascl.soft03023S}, we derive the expected B-V colour for stars with effective temperatures from 50 to 300 kK. To specify, within the PySynphot/Observation module we generated a blackbody spectrum using the BlackBody class, for temperatures 50 to 300 kK (in steps of 5 kK), assuming zero extinction. The B and V synthetic magnitudes in the VEGA system have consequently been obtained for each temperature using the ObsBandpass function. The B-V colour has finally been estimated by subtracting the calculated synthetic B and V magnitudes for each temperature and taking their mean, assigning as error the semi-range of the computed B-V values. Using the mean modelled colour and the b-high - v-high colour of our CSPN, we estimate a CSPN reddening and thus, a PN reddening 
of E(B-V)= 0.92 $\pm$ 0.33 (assuming the PN reddening is identical to the CSPN reddening). We use this PN reddening value in the following analysis, when relevant. Correcting all 
our CSPN magnitudes for interstellar extinction, using the \citet{1989ApJ...345..245C} extinction curve with 
$R_v$=3.1 \citep{1994ApJ...422..158O}, we find u-high = 20.1 $\pm$ 1.1 mag, b-high = 20.7 $\pm$ 1.1 mag, v-high = 
21.1 $\pm$ 1.0 mag, Bessel R special = 21.2 $\pm$ 1.1 mag and Bessel I = 21.1 $\pm$ 1.1 mag.

\section{Analysis \& Discussion} \label{sec:analysis}

The PN systemic radial velocity of -20.0 {\kms} with $\sigma =1.9$ {\kms}, estimated in 
Subsection \ref{subsec:PHR}, is close, within 
the errors, to the cluster's mean radial velocity of -25.5 {\kms} given the unusually large OC velocity dispersion 
for Trumpler~25 that these robust Gaia data provide of $\sigma$=9.1 {\kms}. This provides strong evidence in support 
of PN cluster membership. The PN reddening, as estimated from the CSPN colour (see Subsection \ref{subsec:CSPN}), 
also matches, within the errors providing further support for the physical association of the two objects. 
Taking the mean PN angular radius of 62 arcsec and its integrated $H\alpha$ flux of $logF(H\alpha)=-12.17~mW/m^2$ 
(see Section \ref{sec:intro}) corrected for extinction (from adopting the PN reddening as determined from the CSPN 
colour), we find an extinction corrected $logF(H\alpha)$ of -11.20. We use this to estimate a PN statistical 
distance, using the Surface-Brightness radius (SB-r) relation of \citep{2016MNRAS.455.1459F}, 
of $2194 \pm 500$~pc, which also agrees well with the accurate Gaia based cluster distance of $2417 \pm 14$~pc given the typical $\sim20\%$ error of the SB-r technique. 

Adopting the Gaia cluster mean distance of 2417~pc and the PN mean apparent diameter of 124~arcsec, we estimate a PN 
physical size of 1.78 $\times$ 1.12~pc and a PN mean physical radius of 0.726 $\pm$ 0.004~pc. Considering the derived 
nebular expansion velocity of 31 {\kms}, estimated in Subsection \ref{subsec:PHR}, we obtain a PN kinematic age of 
23000 $\pm$ 6000 years, confirming the highly evolved nature of the PN. The PN kinematic age and so physical 
size and volume, justifies the PN nebular density that for most of our pointings was found to be at the low density limit of $\rm 
N_e < 100$ $\rm cm^{-3}$. From the $H\alpha$/$H\beta$ 
standard ratio of 2.85, we find an extinction corrected nebular $H\beta$ flux of $logF(H\beta)=-11.66~mW/m^2$. 
Following \citet{2015ApJ...803...99C} and using the absolute nebular $H\beta$ flux of -11.66~$mW/m^2$, the electron 
density of $\rm N_e < 100$ $\rm cm^{-3}$ as determined in Subsection \ref{subsec:PHR}, the adopted cluster distance 
of 2417 pc and the effective recombination coefficient of $H\beta$ from \citet{2006agna.book.....O}, we estimate a 
nebular ionized mass larger than 0.11~M$_\sun$. 

The average [\ion{N}{ii}]/H$\alpha$ line ratio, as estimated from the GHOST spectral data, is 10.2, much higher than 
typical for PNe \citep[see,][]{1994MNRAS.271..257K} and indicative of Type-I chemistry 
\citep{1983IAUS..103..233P}. This is also supported by the detection of the He~II $\lambda4686\AA$ high excitation 
emission line, the very high CSPN effective  temperature (see below), and the bipolar nebular shape.

Using our derived CSPN V magnitude, PN reddening (see Subsection~\ref{subsec:CSPN}) and the adopted cluster distance, 
we estimate a CSPN absolute magnitude of $M_v$=9.15 $\pm$ 1.03. The CSPN final mass was derived by over-plotting 
its absolute magnitude and PN kinematic age along with the \citet{2016A&A...588A..25M} evolutionary tracks for 
Z=0.02. We estimate a final mass of $M_{fin}$=0.95 $\pm$ 0.12~M$_\sun$, with the assigned error reflecting the 
absolute magnitude and kinematic age errors.

To estimate the CSPN effective temperature, we over-plot the corresponding luminosities for the estimated 
CSPN absolute magnitude and CSPN effective temperatures from 50 to 300 kK (i.e. the range of possible CSPN temperatures as predicted from theory, see Subsection~\ref{subsec:CSPN}) in steps of 10 kK, along with the 
corresponding \citet{2016A&A...588A..25M} evolutionary tracks \citep[see,][]{2022ApJ...935L..35F}. We derive a 
CSPN effective temperature of 250 $\pm$ 50 kK, which is much higher than average and also indicative of Type~I PN 
chemistry and an intermediate progenitor mass \citep[see,][]{2019NatAs...3..851F}. The assigned temperature error 
reflects the uncertainties associated with the CSPN absolute magnitude. The CSPN effective temperature was also 
estimated following the Zanstra method \citep{1931ZA......2....1Z} and using the absolute nebular $H\beta$ flux, 
the He II $\lambda4686\AA$//H$\beta$ ratio of 1.13 (estimated as an upper limit from our FORS2 data) and the 
CSPN V magnitude, all corrected for extinction, finding a Zanstra $\rm T_{eff}$(HI)=253 $\pm$ 76 kK (adopting 
a 30\% error) and a Zanstra $\rm T_{eff}$(He II) $<$ 290 kK, in excellent agreement with the CSPN effective 
temperature calculated previously from the evolutionary tracks. The estimated CSPN luminosity for an effective 
temperature of 253 $\pm$ 76 kK, is log$ \rm L/L_\sun$= 1.96 $\pm$ 0.55. The estimated physical parameters of 
both the cluster Trumpler~25 and the PN PHR~J1724-3859 are presented in Table~\ref{tab:Tab3}.

\begin{table*}
\caption{Estimated physical properties of the PN PHR~J1724$-$3859 and the open cluster Trumpler~25.}
\centering
\label{tab:Tab3}

\begin{tabular}{lcc}
\hline\hline
\textbf{Parameter} & \textbf{PN PHR J1724$-$3859} & \textbf{Trumpler 25} \\
\hline

%\multicolumn{3}{c}{\textit{Coordinates and Angular/Physical Size}}\\
\hline
RA (J2000)\,(1)     & 17:24:30.70 & 17:24:30.70 \\
DEC (J2000)\,(1)    & $-38$:59:43.91 & $-39$:01:25.00 \\
Apparent size (arcsec$^2$)\,(2),(3) & 152$\times$96 & 576 (diam.) \\
Physical size (pc)\,(4) & 1.78$\times$1.12 & \\
Physical radius (pc)\,(4) & 0.726 $\pm$ 0.004 & 3.37 \\
%\hline

%\multicolumn{3}{c}{\textit{Kinematics, Distance, and Reddening}}\\
%\hline
Radial velocity ({\kms})\,(5),(6) & $-20.0$, $\sigma=1.9$ & $-25.5$, $\sigma=9.1$ \\
Distance (kpc)\,(7),(8) & $2.2 \pm 0.5$ & $2.4 \pm 0.01$ \\
Reddening $E(B-V)$\,(9),(10) & $0.92 \pm 0.33$ & $0.85 \pm 0.01$ \\
%\hline

%\multicolumn{3}{c}{\textit{Nebular Properties}}\\
%\hline
Expansion velocity ({\kms})\,(11) & 31, $\sigma=8$ & \\
Chemistry\,(12)      & Type~I & [Fe/H] $=0.17 \pm 0.02$ \\
Ionised mass (M$_\sun$)\,(15) & $>$0.11 & \\
Electron density $N_e$ (cm$^{-3}$)\,(16) & $<$100 & \\
log(H$\alpha$) flux ($mW/m^2$)\,(17) & $-12.17 \pm 0.12$ & \\
%\hline

%\multicolumn{3}{c}{\textit{Central Star Parameters}}\\
%\hline
$T_{\rm eff}$ (kK)\,(18) & $250 \pm 50$ & \\
Zanstra T(H I) (kK)\,(19) & $253 \pm 76$ & \\
Zanstra T(He II) (kK)\,(20) & $<$290 & \\
Luminosity log$L/L_\sun$\,(21) & $1.96 \pm 0.55$ & \\
Initial mass (M$_\sun$) & $5.12^{+0.16}_{-0.15}$ & \\
Final mass (M$_\sun$)   & $0.95 \pm 0.12$ & \\
%\hline

%\multicolumn{3}{c}{\textit{Photometry and Proper Motions}}\\
%\hline
pmRA (mas\,yr$^{-1}$)\,(22)  & & $0.315 \pm 0.002$ \\
pmDec (mas\,yr$^{-1}$)\,(22) & & $-2.138 \pm 0.002$ \\
CSPN magnitudes (U,B,V,R,I)\,(23) & 20.1–21.2 & \\
Absolute magnitude $M_V$\,(24) & $9.15 \pm 1.03$ & \\
%\hline

%\multicolumn{3}{c}{\textit{Age}}\\
%\hline
Age\,(13),(14) & $23 \pm 6$ kyr & $112 \pm 9$ Myr \\
\hline

\end{tabular}

\tablecomments{
(1) HASH database \citep{2016JPhCS.728c2008P}.  
(2) PN apparent size from \citet{2006MNRAS.373...79P}.  
(3) Cluster diameter from \citet{2013A&A...558A..53K}.  
(4) Computed for the adopted cluster distance.  
(5) PN radial velocity from GHOST+HRS data.  
(6) Cluster RV from Gaia DR3 \citep{2023A&A...675A..68V}.  
(7) PN statistical distance \citep{2016MNRAS.455.1459F}.  
(8) Cluster distance from Gaia parallaxes \citep{2023A&A...675A..68V}.  
(9) PN reddening from CSPN colour.  
(10) Cluster reddening from AsteCa isochrone fit.  
(11) Expansion velocity from GHOST/HRS line splitting and H$\alpha$ FWHM.  
(12) Cluster metallicity from AsteCa isochrones ($Z=0.023\pm0.001$).  
(13) PN age from radius, velocity, and distance.  
(14) Cluster age from Padova isochrones.  
(15) Nebular mass from \citet{2015ApJ...803...99C}.  
(16) Electron density from [S II] ratio assuming $T_e=10^4$\,K.  
(17) H$\alpha$ flux from \citet{2013MNRAS.431....2F}.  
(18) Derived from \citet{2016A&A...588A..25M} tracks.  
(19) Zanstra temperature (H I).  
(20) Upper limit from He II $\lambda4686$/H$\beta$.  
(21) Luminosity using Zanstra T(HI).  
(22) Gaia DR3 proper motions.  
(23) CSPN magnitudes from FORS2 photometry.  
(24) Absolute magnitude from cluster distance and PN reddening.  
}

\end{table*}

\begin{figure*}
\plotone{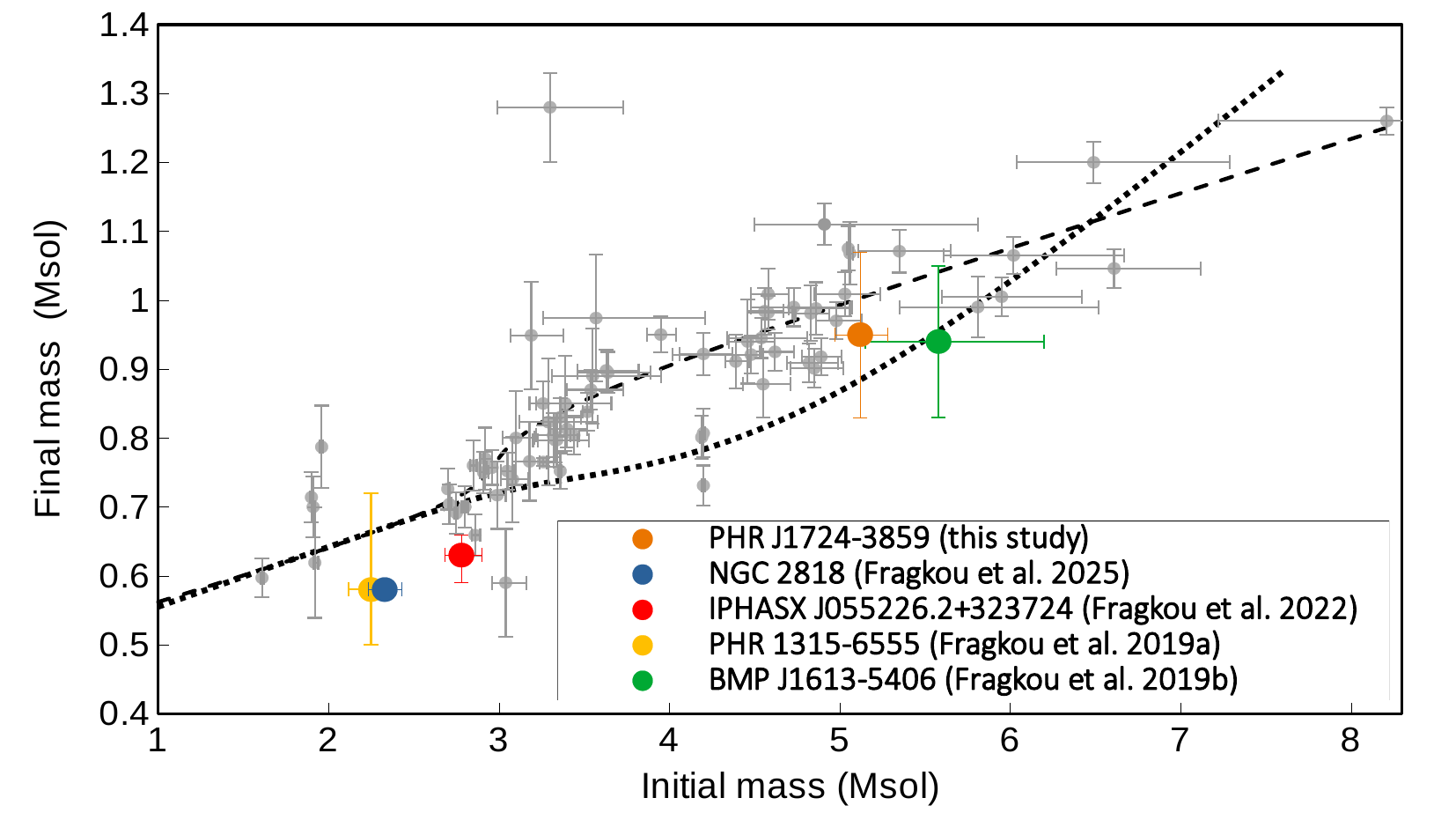}
\caption{Our open cluster PN points (coloured filled circles) over-plotted along the 
IFMR "PARSEC" 
trend (dashed line) of the  \citet{2018ApJ...866...21C} data points (grey filled circles). The latest IFMR estimated by \citet{2024MNRAS.527.3602C} using population synthesis modeling is shown by the dotted line.}
\label{Fig9}
\end{figure*}

The initial-to-final mass relation (IFMR) is traditionally constructed from cluster white dwarfs \citep[e.g.,] [] 
{2018ApJ...866...21C} and links their physical properties to those of their progenitor stars, determined 
independently from cluster studies. Cluster PNe can offer additional key data points to contribute in the 
establishment of a distinct IFMR that is currently poorly constrained, which is crucial in tracing the 
enrichment of essential elements, such as carbon and nitrogen, in our Galaxy and beyond. Figure~\ref{Fig9} 
shows our new cluster PN point (i.e. the initial and final masses of the CSPN of PHR~J1724-3859 estimated above) 
and all other points derived from our other PNe associated with open clusters 
\citep{2019MNRAS.484.3078F,2019NatAs...3..851F,2022ApJ...935L..35F} over-plotted along with the 
latest IFMR estimates from cluster WDs and semi-empirical “PARSEC” fit \citep{2018ApJ...866...21C} and the \citet{2024MNRAS.527.3602C} IFMR derived from a volume-limited 40 pc sample of WDs using population synthesis modelling. Our new data point, 
depicted in orange, falls in the sparsely populated intermediate initial mass area of the IFMR,  
and agrees, within the errors, with both the predictions from cluster white dwarfs and population synthesis models, and the other 
open cluster PN points (see Table~\ref{tab:Tab4}).  All open cluster-PN points fall 
slightly lower, by $\sim0.1M_\sun$, than the general trend from \citet{2018ApJ...866...21C}, a systematic 
effect that deserves to be further explored in future studies and possibly associated with the evolutionary tracks 
employed for the estimation of final masses. The \citet{2024MNRAS.527.3602C} IFMR seems to better fit our data for intermediate initial masses (with our new cluster PN point falling above their trend) but unlike an empirical or semi-empirical IFMR, derived from cluster WDs \citeg{2018ApJ...866...21C}, it does not implicity include the effects of stellar rotation and magnetic fields that could affect the final masses of mainly intermediate and higher mass stars.

\begin{table*}[]
\caption{The initial and final masses of all open cluster PNe.}
\centering
\label{tab:Tab4}
\begin{tabular}{lll}
\hline\hline
PN                           & CSPN $M_{ini}$ ($M_\sun$) & CSPN $M_{fin}$ ($M_\sun$) \\
\hline
PHR~J1724-3859          & 5.12$_{-0.15}^{+0.16}$    & 0.95$\pm$0.12             \\
PHR~1315-6555\footnote{From \citet{2019MNRAS.484.3078F}.}           & 2.25$\pm$0.13             & 0.58$_{-0.08}^{+0.14}$    \\
BMP~J1613-5406\footnote{From \citet{2019NatAs...3..851F}.}              & 5.58$_{-0.43}^{+0.62}$    & 0.94$\pm$0.11             \\
IPHASX~J055226.2+323724\footnote{From \citet{2022ApJ...935L..35F}.}     & 2.78$_{-0.10}^{+0.12}$    & 0.63$_{-0.04}^{+0.03}$    \\
NGC~2818\footnote{From \citet{Fragkou2025}.}                    & 2.33$\pm$0.10             & 0.58$\pm$0.01       \\  
\hline
\end{tabular}
\end{table*}

\section{Conclusions}\label{sec:conc}

PNe that are physical members of star clusters are very rare, but due to their importance in stellar 
evolution studies, any additional detection is highly significant. 

Using data acquired from a series of large telescopes we identified the CSPN of the proposed new OC-PN pair
and estimated its physical properties along with the properties of its progenitor, the PN PHR~J1724-3859 and 
the host cluster Trumpler~25.

Using our high-resolution spectra we showed that the PN PHR~J1724-3859 has a mean systemic radial velocity 
of -20.0 {\kms} with $\sigma$=1.9  {\kms} very close to that of the open cluster Trumpler~25 which has been 
estimated to be -25.5 {\kms} with $\sigma$= 9.1 {\kms} from Gaia DR3 data. This is key evidence supporting 
the physical association of the two objects  given the cluster has an unusually large, but robustly 
determined, velocity dispersion. Then taking into account the good agreement of the estimated OC-PN 
reddening and the independently estimated PN and cluster distances, the proximity of the PNe to the 
cluster core (well within the tidal radius), and plausible derived PN physical parameters, 
PN cluster membership appears to be highly likely. When the properties of the PN in this new OC-PN pair 
are also further compared to the four other members of the class, there are striking common properties. 
All are bipolars, all have Type~I chemistry shown by nitrogen enrichments with large H$\alpha$ to [NII] ratios, 
all are kinematically at extreme limits and all have very hot CSPN. These combined properties are 
themselves rare amongst the general PNe population 
(see Table~3 in \cite{Fragkou2025}). They are also evolved and have higher than typical progenitor masses. 
This is explained by the fact that, given the cluster ages, only high mass progenitors can currently form PNe 
in such clusters. PNe with low-mass cluster star progenitors are still billions of years away from forming PNe. 

Therefore, we conclude that the PHR~J1724-3859-Trumpler~25 pair is highly likely to be a real association and 
include it in our small sample of very rare open cluster-PNe associations.

Our findings again are in mild disagreement with the latest IFMR estimates from cluster WDs from \citet{2018ApJ...866...21C} 
but fit the emerging trend for all our other open cluster-PNe. They also better agree with the \citet{2024MNRAS.527.3602C} IFMR from population synthesis models. We are undertaking studies to explore the origin of the 
systematic effect observed in the final masses of open cluster-PNe, which fall slightly below by $\sim0.1M_\sun$ 
but effectively parallel to the general trend of cluster white dwarfs (Parker et al., in  preparation). 
This could have implications for stellar evolutionary timescales.

%\section{Software and third party data repository citations} \label{sec:cite}

%Authors can also use the website \url{https://www.doi2bib.org/} to create a BIBTeX entry for any DOI. Please check the output from this site carefully as its output is only as good as the DOI metadata. Some DOI creators do not provide enough metadata to construct an adequate citation.

%% Please use the acknowledgment and contribution environments. This will 
%% be anonomyized when the "anonymous" style option is used. 

\begin{acknowledgments}
VF thanks Fundação Coordenação de Aperfeiçoamento de Pessoal de Nível Superior (CAPES) for granting the postdoctoral research fellowship Programa de Desenvolvimento da Pós-Graduação (PDPG)-Pós Doutorado Estratégico, Edital nº16/2022, Coordenação de Aperfeiçoamento de Pessoal de Nível Superior - Brasil (CAPES) – Finance Code 001 (88887.838401/2023-00) and Funda\c{c}\~ao de Amparo \`a Pesquisa do Estado do Rio de Janeiro (FAPERJ) for granting the postdoctoral research fellowships E-26/200.181/2025 and 200.181/2025(304980). QAP thanks the Hong Kong Research Grants Council for GRF research grants 17326116, 17300417 and 17304520. DRG acknowledges FAPERJ (E-26/211.527/2023) and CNPq (315307/2023-4) for partical support.

We made use of NASA’s Astrophysics Data System; the SIMBAD database operated at CDS, Strasbourg, France; Astropy, a community-developed core Python package for Astronomy (T. A. Collaboration et al. 2022); HASH, an online database at the Laboratory for Space Research at HKU federates available multi-wavelength imaging, spectroscopic, and other data for all known Galactic PNe and is available at: http://www.hashpn.space. 

Some of the observations reported in this paper were obtained with the Southern African Large Telescope (SALT) under program 2019-1-DDT-002 (PI: Q. A. Parker).  Others were  based on observations collected at the European Southern Observatory under ESO programms 0103.D-0093(A) and 0103.D-0093(B). More observations were obtained at the international Gemini Observatory, a program (Program ID: GS-2024A-Q-232) of NSF NOIRLab (acquired through the Gemini Observatory Archive), at NSF NOIRLab and processed using the Gemini IRAF package and DRAGONS (Data Reduction for Astronomy from Gemini Observatory North and South), which is managed by the Association of Universities for Research in Astronomy (AURA) under a cooperative agreement with the U.S. National Science Foundation on behalf of the Gemini Observatory partnership: the U.S. National Science Foundation (United States), National Research Council (Canada), Agencia Nacional de Investigaci\'{o}n y Desarrollo (Chile), Ministerio de Ciencia, Tecnolog\'{i}a e Innovaci\'{o}n (Argentina), Minist\'{e}rio da Ci\^{e}ncia, Tecnologia, Inova\c{c}\~{o}es e Comunica\c{c}\~{o}es (Brazil), and Korea Astronomy and Space Science Institute (Republic of Korea).

This work made use of the University of 
Hong Kong/Australian Astronomical Observatory/Strasbourg Observatory H-alpha Planetary Nebula (HASH PN) database, hosted by the Laboratory for Space Research at the University of Hong Kong; This research has made use of the SIMBAD database and the VizieR catalogue access tool, CDS, Strasbourg, France (doi: 10.26093/cds/vizier). The original description of the VizieR service was published in Ochsenbein, Bauer \& Marcout (2000).

This work has made use of data from the European Space Agency (ESA) mission Gaia (https://www.cosmos.esa.int/gaia), processed by the Gaia Data Processing and Analysis Consortium (DPAC, 
https://www.cosmos.esa.int/web/gaia/dpac/consortium).
\end{acknowledgments}

\vspace{5mm}
\facilities{SALT (HRS), VLT (FORS2), Gemini (GHOST)}

\software{HRS pipeline \citep{2016MNRAS.459.3068K,2017ASPC..510..480K},  
          EsoReflex \citep{2013A&A...559A..96F}, 
          DRAGONS \citep{2023RNAAS...7..214L},
          Source Extractor \citep{1996A&AS..117..393B}
          ASteCA \citep{2015A&A...576A...6P}
          }
%GALAXEV, \citep{Bruzual:2003}, GMM \citep{Muratov:2010}, GTCMOS \citep{Mauricio:2016}, IspecFit \citep{Lomeli:2024}.
%astropy \citep{2013A&A...558A..33A,2018AJ....156..123A,2022ApJ...935..167A},  
%          Cloudy \citep{2013RMxAA..49..137F}, 
%          Source Extractor \citep{1996A&AS..117..393B}


%\appendix

%% For this sample we use BibTeX plus aasjournalv7.bst to generate the

%\bibliography{sample7}{}
%\bibliography{example}{}
\bibliography{sample631}{}
\bibliographystyle{aasjournalv7}



\end{document}